\documentclass[sigplan,10pt,screen]{acmart}

\usepackage{microtype}
\usepackage[utf8]{inputenc}
\usepackage[T1]{fontenc}

\usepackage{listings}
\usepackage{xspace}
\usepackage{algorithm}
\usepackage{algorithmic}
\usepackage[english]{babel}
\usepackage{multirow}
\usepackage{color}
\usepackage{colortbl}
\usepackage{tabularx}
\usepackage{xcolor}
\usepackage{multirow}
\usepackage{subfigure}
\usepackage{eqparbox}
\usepackage{natbib}
\usepackage{doi}

\usepackage{etoolbox}  
\makeatletter
\patchcmd{\algorithmic}{\addtolength{\ALC@tlm}{\leftmargin} }{\addtolength{\ALC@tlm}{\leftmargin}}{}{}
\def\@copyrightspace{\relax}
\makeatother

\theoremstyle{definition}
\newtheorem{definition}{Definition}[section]

\AtBeginDocument{%
  \providecommand\BibTeX{{%
    \normalfont B\kern-0.5em{\scshape i\kern-0.25em b}\kern-0.8em\TeX}}}

\lstdefinelanguage{JavaScript}{
  keywords={typeof, new, true, false, catch, function, return, null, catch, switch, var, if, in, while, do, else, case, break},
  keywordstyle=\color{blue}\bfseries,
  ndkeywords={class, export, boolean, throw, implements, import, this},
  ndkeywordstyle=\color{darkgray}\bfseries,
  identifierstyle=\color{black},
  sensitive=false,
  comment=[l]{//},
  morecomment=[s]{/*}{*/},
  commentstyle=\color{purple}\ttfamily,
  stringstyle=\color{red}\ttfamily,
  morestring=[b]',
  morestring=[b]"
}

\colorlet{punct}{red!60!black}
\definecolor{background}{HTML}{EEEEEE}
\definecolor{delim}{RGB}{20,105,176}
\colorlet{numb}{magenta!60!black}

\lstdefinelanguage{json}{
    basicstyle=\normalfont\ttfamily,
    numbers=left,
    numberstyle=\scriptsize,
    stepnumber=1,
    numbersep=8pt,
    showstringspaces=false,
    breaklines=true,
    frame=lines,
    literate=
     *{0}{{{\color{numb}0}}}{1}
      {1}{{{\color{numb}1}}}{1}
      {2}{{{\color{numb}2}}}{1}
      {3}{{{\color{numb}3}}}{1}
      {4}{{{\color{numb}4}}}{1}
      {5}{{{\color{numb}5}}}{1}
      {6}{{{\color{numb}6}}}{1}
      {7}{{{\color{numb}7}}}{1}
      {8}{{{\color{numb}8}}}{1}
      {9}{{{\color{numb}9}}}{1}
      {:}{{{\color{punct}{:}}}}{1}
      {,}{{{\color{punct}{,}}}}{1}
      {\{}{{{\color{delim}{\{}}}}{1}
      {\}}{{{\color{delim}{\}}}}}{1}
      {[}{{{\color{delim}{[}}}}{1}
      {]}{{{\color{delim}{]}}}}{1},
}

\newcommand\JS{\texttt{JS}\xspace}

\newcommand\cparagraph[1]{\vspace{1mm}\noindent\textbf{#1}\xspace}
\newcommand{\ecma}{ECMA-262\xspace}

\newcommand{\SystemName}{\textsc{Comfort}\xspace}

\newcommand{\SBugs}{158\xspace}
\newcommand{\VBugs}{129\xspace}
\newcommand{\FBugs}{115\xspace}
\newcommand{\TSC}{21\xspace}

\definecolor{Gray}{gray}{0.95}
\definecolor{highlight}{rgb}{1,1,0.04}

\setcopyright{acmcopyright}
\acmPrice{15.00}
\acmDOI{10.1145/3453483.3454054}
\acmYear{2021}
\copyrightyear{2021}
\acmSubmissionID{pldi21main-p149-p}
\acmISBN{978-1-4503-8391-2/21/06}
\acmConference[PLDI '21]{Proceedings of the 42nd ACM SIGPLAN International Conference on Programming Language Design and Implementation}{June 20--25, 2021}{Virtual, Canada}
\acmBooktitle{Proceedings of the 42nd ACM SIGPLAN International Conference on Programming Language Design and Implementation (PLDI '21), June 20--25, 2021, Virtual, Canada}

\begin{document}

\title{Automated Conformance Testing for JavaScript Engines via Deep Compiler Fuzzing}

\author{Guixin Ye}
\email{gxye@nwu.edu.cn}
\affiliation{
    \institution{Northwest University} 
    \city{Xi'an}
    \country{China}
}

\author{Zhanyong Tang}
\email{zytang@nwu.edu.cn}
\affiliation{
    \institution{Northwest University} 
    \city{Xi'an}
    \country{China}
}

\author{Shin Hwei Tan}
\email{tansh3@sustech.edu.cn}
\affiliation{
    \institution{Southern University of Science and Technology} 
    \city{Shenzhen}
    \country{China}
}

\author{Songfang Huang}
\email{songfang.hsf@alibaba-inc.com}
\affiliation{
    \institution{Alibaba DAMO Academy}
    \city{Beijing}
    \country{China}
}

\author{Dingyi Fang}
\email{dyf@nwu.edu.cn}
\affiliation{
    \institution{Northwest University} 
    \city{Xi'an}
    \country{China}
}

\author{Xiaoyang Sun}
\email{scxs@leeds.ac.uk}
\affiliation{
    \institution{University of Leeds} 
    \city{Leeds}
    \country{United Kingdom}
}

\author{Lizhong Bian}
\affiliation{
    \institution{Alipay (Hangzhou) Information \& Technology Co., Ltd.} 
    \city{Hangzhou}
    \country{China}
}

\author{Haibo Wang}
\email{schw@leeds.ac.uk}
\affiliation{
    \institution{University of Leeds} 
    \city{Leeds}
    \country{United Kingdom}
}

\author{Zheng Wang}
\email{z.wang5@leeds.ac.uk}
\affiliation{
    \institution{University of Leeds} 
    \city{Leeds}
    \country{United Kingdom}
}

\thanks{Corresponding authors: Zhanyong Tang, Songfang Huang and Zheng Wang.}

\renewcommand{\shortauthors} {Guixin Ye, et al.}

\begin{abstract}
JavaScript (\JS) is a popular, platform-independent programming language. To ensure the interoperability of \JS programs across different
platforms, the implementation of a \JS engine should conform to the ECMAScript standard. However, doing so is challenging as there are
many subtle definitions of API behaviors, and the definitions keep evolving.

We present \SystemName, a new compiler fuzzing framework for detecting \JS engine bugs and behaviors that deviate from the ECMAScript
standard. \SystemName leverages the recent advance in deep learning-based language models to automatically generate \JS test code. As a
departure from prior fuzzers, \SystemName utilizes the well-structured ECMAScript specifications to automatically generate test data
along with the test programs to expose bugs that could be overlooked by the developers or manually written test cases. \SystemName then
applies differential testing methodologies on the  generated test cases to expose standard conformance bugs. We apply \SystemName to ten
mainstream \JS engines. In 200 hours of automated concurrent testing runs, we discover bugs in all tested \JS engines. We had identified
\SBugs unique \JS engine bugs, of which \VBugs have been verified, and \FBugs have already been fixed by the developers. Furthermore,
\TSC of the \SystemName-generated test cases have been added to Test262, the official ECMAScript conformance test suite.

\end{abstract}

\ccsdesc[500]{Software and its engineering~Compilers}
\ccsdesc[500]{Software and its engineering~Language features}
\ccsdesc[300]{Computing methodologies~Artificial intelligence}

\keywords{JavaScript, Conformance bugs, Compiler fuzzing, Differential testing, Deep learning}

\maketitle

\section{Introduction}
JavaScript (\JS) is one of the most popular programming languages \cite{githubc}. It is a core technology that underpins the web browser,
server-side web deployments and many embedded and smartphone applications. The implementation of a \JS engine (compiler) should conform to
the ECMAScript specification, ECMA-262~\cite{ecmascript2020}, that ensures the interoperability of \JS code across different \JS platforms.
Non-standard \JS engine implementations can confuse the application developer, leading to unexpected software behavior and poor user
experience during deployment. Still, writing a \JS engine that conforms to ECMAScript is hard due to the complexity of modern \JS
interpreters, the large number of \JS APIs\footnote{The ECMA-262 2020 Edition defines over 750 \JS APIs.} and object types, the constantly
evolving language specification, and the diversity of \JS code and inputs seen in real-life deployments. Hand-written \JS test cases for
testing \JS engines, while important, are inadequate for covering all parts of the \JS language standard to test a sophisticated \JS
engine.

Random test case generation - or \emph{fuzzing} - is a widely used technique for automated compiler bug detection
\cite{chen2013taming,chen2016empirical,manes2019the}. It is often used together with differential testing
\cite{pldi14,sun2016finding,ofenbeck2016randir} to discover unexpected program behavior. In the context of fuzzing \JS compilers,  a
randomly generated \JS program and its input forms a \emph{test case}, which is executed by \emph{multiple} \JS engines. Any unexpected
behavior, including crashing, freezing or inconsistent compliation or execution outcomes among engines, indicates a \JS compiler bug.

The success of compiler fuzzing requires generating bug-exposing test cases with the right program and test input to trigger buggy
behavior~\cite{chen2020survey}. Unfortunately, doing so is challenging as there are many ways to construct a program and its input.
Existing fuzzing techniques for \JS typically follow a generative or mutational approach. Generative approaches build a new test case from
the ground up using predefined grammar rules \cite{yang2011finding, holler2012fuzzing} or by reassembling synthesizable code segments from
a program corpus \cite{park2020fuzzing,han2019codealchemist}. By contrast, mutational approaches synthesize a test case from existing seed
programs and inputs \cite{wang2019superion}. Both strategies require expert involvement to construct the grammar rules or preparing a
high-quality seed code corpus to ensure the coverage of the test cases, but doing so becomes increasingly difficult due to the complexity
and constantly evolving nature of the \JS language standard.

We present \SystemName\footnote{\SystemName = \textbf{COM}piler \textbf{F}uzzing f\textbf{O}r javasc\textbf{R}ip\textbf{T}}, a
\emph{generative} fuzzer for \JS engines. Unlike existing \JS fuzzers  that aim to detect crashing bugs or vulnerabilities
\cite{park2020fuzzing}, \SystemName focuses on exposing standard conformance bugs. \SystemName leverages the advances in deep-learning (DL)
based program synthesis \cite{cummins2018compiler} to generate \JS programs by automatically learning a generation model. Specifically,
\SystemName employs GPT-2~\cite{radford2019language}, a recently proposed language generation model, to generate \JS code by learning from
a corpus of open-source \JS programs. GPT-2 improves the long short-term memory (LSTM) model used in state-of-the-art DL-based fuzzers
\cite{cummins2018compiler,liu2019deepfuzz,son2020montage} by generating valid \JS programs with a higher success rate, as it can model
longer dependencies in the program source code. \SystemName then uses differential testing to detect buggy \JS engine behavior.

Existing compiler fuzzers use a random input generation strategy by relying on the typing information of a variable \cite{pldi14,
cummins2018compiler} to generate kernel or function parameters. However, \JS is a weakly typed language, where a variable can be of an
arbitrary type and can be of multiple types throughout the execution. This feature increases the space of possible input settings, making
it harder for a random input generation strategy to trigger compiler bugs with reasonable cost. To effectively generate test program
inputs, \SystemName draws hints from ECMA-262. It leverages the well-structured specification rules defined in the specification document
to narrow down the scope of argument types and boundary values for \JS APIs and edge cases that are likely to trigger unusual behavior. By
narrowing down the scope, \SystemName is able to generate inputs that are more likely to cover the special cases overlooked by \JS engine
developers and manually written test cases. The result is a new way of leveraging the language specification for compiler fuzzing.

We evaluate \SystemName\footnote{Code and data are available at: \url{https://github.com/NWU-NISL-Fuzzing/COMFORT}.} by applying it to ten
mainstream \JS engines. The \JS engines that we target include those used in mainstream web browsers: JavaScriptCore (JSC) in Apple Safari, V8 in
Google Chrome, ChakraCore in Microsoft Edge, SpiderMonkey in Firefox, \JS engines for mobile and embedded systems: Hermes, QuickJS, Rhino,
Nashorn, and JerryScript, and Graaljs that is specifically designed to be compatible with ECMAScript 2020. Our large-scale evaluation shows
that \SystemName is highly effective in generating syntactically correct \JS programs with a better test coverage, where 80\% of the
generated code is syntactically correct. This success rate translates to $2.6\times$ improvement over the success rate given by DeepSmith
\cite{cummins2018compiler}, a state-of-the-art DL-based generative fuzzer. We show that \SystemName is more efficient in producing
bug-exposing test cases by uncovering at least $2\times$ more unique bugs within the same test running time, when compared to
state-of-the-art \JS fuzzers \cite{han2019codealchemist,son2020montage,gross2018fuzzil,park2020fuzzing}. In 200 hours of automated
concurrent testing runs, \SystemName discovers bugs in all tested \JS engines. We have identified \SBugs unique \JS compiler bugs, covering
109 newly discovered bugs. Of the submitted bugs, \VBugs have been verified and \FBugs have been fixed by the relevant \JS compiler
developers. Moreover, \TSC of the test cases produced by \SystemName has been added to Test262~\cite{test262}, the official ECMAScript conformance test suite.

This paper shares our experience and findings of exploiting ECMA-262 to detect \JS compiler bugs through fuzzing. It makes the following
contributions:

\begin{itemize}
    \item It is among the first studies on employing compiler fuzzing to expose conformance bugs in \JS compilers;
    \item It is the first random program generator for leveraging the language specification document to generate test data for compiler
    fuzzing (Section \ref{sec:tdg});
    \item It provides a large study independently validating the effectiveness of the recently proposed DL-based test program generation
        method \cite{cummins2018compiler} in a new domain (Section \ref{sec:generation}).
\end{itemize}
\section{Background and Motivation}
\begin{figure}[t!]
    \centering
      \includegraphics[width=0.45\textwidth]{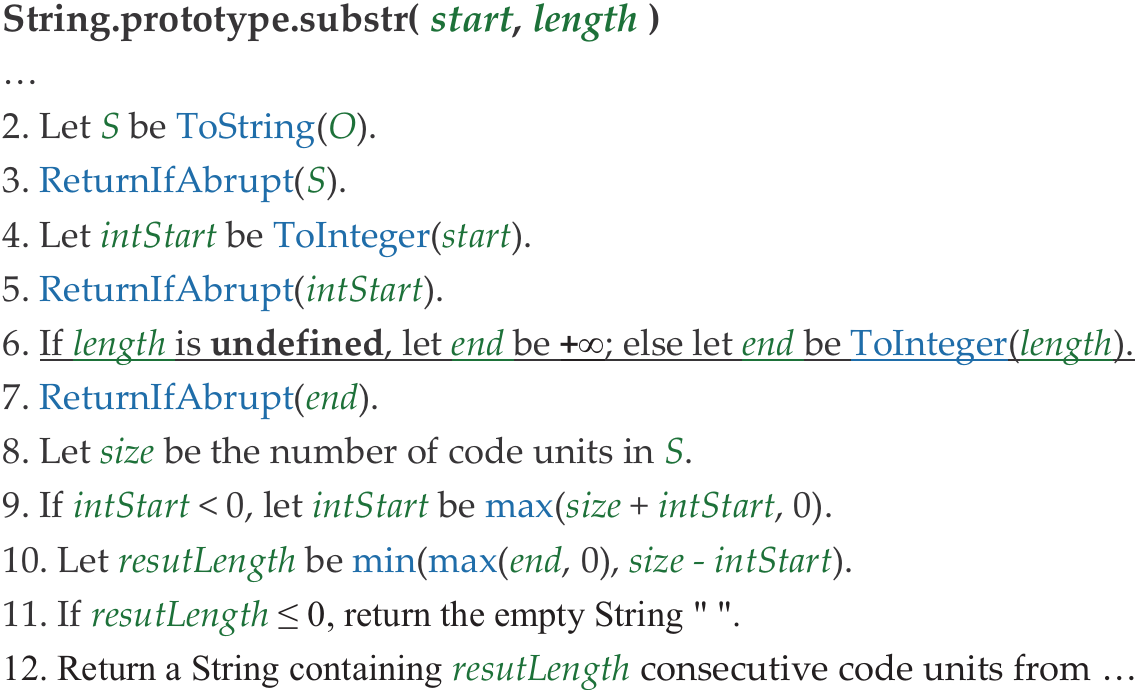}
      \vspace{-2mm}
    \caption{Expected behavior of the \texttt{substr} function defined in the ECMAScript 2015 specification.}
    \label{fig:pseudocode}
    \vspace{-3.5mm}
\end{figure}

\subsection{JavaScript Standard}
ECMA-262\footnote{Another ECMAScript document called ECMA-402 provides additional definitions for internationalization APIs for, e.g., the
numbering system, calendar, and formats, used by different human languages and countries. We do not target ECMA-402 in this work.} is the
standardized specification for JavaScript. The first edition of \ecma was developed in 1997, and the 11th edition, officially known as
ECMAScript 2020, was published in June 2020. ECMA-262 is a well-structured document that uses a mixture of natural language descriptions
and pseudo-code to describe the expected observed behavior for \JS APIs. Figure \ref{fig:pseudocode} shows an example pseudo-code for the
\texttt{substr} function defined in ECMA-262.

Test262 \cite{test262} is an ECMA-262 conformance test suite for checking how closely a \JS implementation follows the ECMAScript
specification. As of August 2020, this test suite has 140 contributors, containing over 38,000 test cases - each of which tests some
specific ECMA-262 specification rules.

\vspace{-2mm}
\subsection{Problem Scope}
We now define the notion of the \emph{conformance bug} for the \JS engine, and then introduce the concept of the \emph{conformance testing} before scoping our work.

\theoremstyle{definition}
\begin{definition}[Conformance bug]
Given implementations of multiple \JS engines $\mathbb{J}$ and a version of the \ecma{} specification $\mathbb{E}$, a \emph{conformance bug} is an unexpected behavior in $\mathbb{J}$ that occurs due to a violation to the specification in $\mathbb{E}$.
\end{definition}

\theoremstyle{definition}
\begin{definition}[Conformance testing]
Given an implementation $\tau_i$ of a \JS engine and an \ecma{} specification $\mathbb{E}$, \emph{conformance testing} is a technique that
aims to identify if $\tau_i$ meets the requirements of $\mathbb{E}$. Conformance testing is used to discover conformance bugs.
\end{definition}

Prior work on \JS compiler fuzzing is primarily concerned about detecting crashing bugs or
vulnerabilities~\cite{han2019codealchemist,son2020montage,gross2018fuzzil,park2020fuzzing}, but detecting engine conformance using fuzzing
is largely overlooked. \SystemName is designed to expose JS compiler bugs, including conformance bugs. We remark that \SystemName is not
designed to automate bug confirmation. Instead, it will report the potential buggy behavior to the JS engine developers to ask for
developer confirmations. In this work, when testing a \JS engine, we restrict our scope to the ECMA-262 edition that the targeting engine
version claims to be compatible with. Therefore, we remove test programs that contain unsupported JS APIs in our test dataset when fuzzing
a \JS engine. We also manually inspected the failed test cases of a compiler version and only reported bugs if the feature is supported by
the relevant \ecma version.

\begin{figure}[!t]
    \centering
\begin{lstlisting}[basicstyle = \scriptsize\ttfamily]
function foo(str, start, len) {
    var ret = str.substr(start, len); 
    return ret;
}
var s = "Name: Albert";
var pre = "Name: ";
var len = undefined;
var name = foo(s, pre.length, len);
print(name);
\end{lstlisting}
    \vspace{-3mm}
    \caption{A test case generated by \SystemName, which exposes a conformance bug of Rhino.
    }
    \label{fig:motivation_example}
    \vspace{-4mm}
\end{figure}

\subsection{Using \JS Specifications for Fuzzing}
\vspace{-1mm} To demonstrate how ECMA-262 can help in identifying conformance bugs, consider the \JS example shown in
Figure~\ref{fig:motivation_example}.

This code uses the \texttt{substr} API (line 2) to extract a substring of \texttt{str}, starting at the position defined by the
\texttt{start} argument and  extending for a given number (defined by \texttt{len}) of characters afterwards. According to the \ecma rules
in Figure \ref{fig:pseudocode} (line 6), if the parameter \texttt{length} (\texttt{len} in Figure~\ref{fig:motivation_example}) is
\emph{undefined}, the function should extract the substring from \texttt{start} to the end of the string. For this example program, it
should print ``Albert'' as an output. However, the latest version of the Rhino \JS engine produces an empty string, which is thus a
standard conformance bug. To generate this bug-exposing input, the fuzzer needs to be aware of the context, i.e., it needs to produce a
String object and ensure the \texttt{len} variable (line 7 in Figure \ref{fig:pseudocode}) is \emph{undefined} before passing to
\texttt{substr}. Existing compiler fuzzers would struggle to generate such a test case because they typically use an input generation
strategy to assign random values to variables; since simply leaving a variable undefined before it is used without knowing the context will
frequently trigger many correctly handled runtime exceptions. As we will show later in the paper,  by using ECMA-262 to guide the test
input generation, \SystemName successfully produced this test case and uncovered a new bug in Rhino, which was not covered by the
hand-written Test262 test suite.

\begin{figure*}
    \centering
    \includegraphics[width=0.95\textwidth]{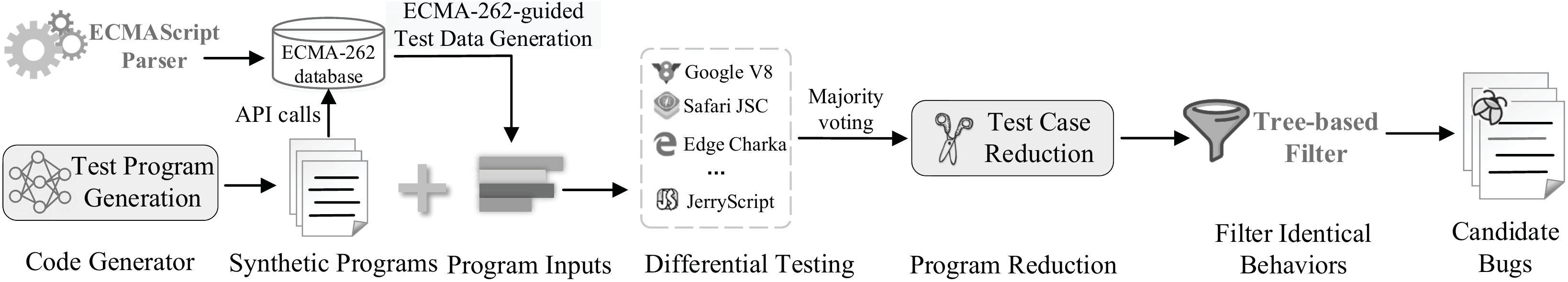}
    \vspace{-2mm}
    \caption{Overview of \SystemName. We use GPT-2 to generate test \JS programs. For \JS APIs in a test program, we utilize the API
    definitions and boundary conditions from the \ecma database to generate test data. We execute multiple \JS engines on the
    test cases and use differential testing to identify potential buggy behaviors.}
    \label{fig:overview}
    \vspace{-2mm}
\end{figure*}

\vspace{-2mm}
\subsection{Transformer-based Test Program Generation}
\vspace{-1mm} Our program generator is based on GPT-2, a Transformer-based neural generation architecture \cite{radford2019language}. Most
of the recently proposed deep-learning-based compiler fuzzers
\cite{son2020montage,cummins2018compiler,godefroid2017learn,he2019learning,liu2019deepfuzz} use a recurrent neural network (RNN), e.g.,
LSTM \cite{hochreiter1997long}, to model the program source code sequence to generate test programs. A Transformer architecture does not
rely on any RNN mechanism. Instead, it uses an attention mechanism to model the sequence dependence. The attention mechanism takes an input
sequence and decides at each step which other parts of the sequence are important. Prior studies showed that the Transformer outperforms
RNN in modeling long-term sequence dependence for natural language processing~\cite{merkx2020comparing}. In this work, we utilize an
open-source pre-trained GPT-2 model
\cite{radford2019language} but re-target it to model and generate \JS programs (Section \ref {sec:generation}).

\section{\SystemName}
Figure~\ref{fig:overview} provides a high-level overview of \SystemName. \SystemName is designed to use the language specification to guide
test case generation. To this end, we build an automated parser to extract the pseudo-code-like \JS API rules (see
Figure~\ref{fig:pseudocode} for an example) from ECMA-262. We store the parsing results in a structured database. To generate \JS
\emph{test programs}, we first use our GPT-2 program generator to produce random \JS programs (Section \ref{sec:generation}). To create
\emph{test data} for a test program, \SystemName first extracts the APIs and their arguments from the program (Section \ref{sec:tdg}). It
then looks up the extracted API rules to generate inputs (i.e., by assigning values to variables in the \JS code) that can trigger the
boundary conditions of an API definition. To enrich the pool of inputs, we also generate some random input values. A \JS test program and
one of its datasets then form a \emph{test case}, and a test program can be associated with multiple input datasets. The generated test
cases are used to test \JS engines through differential testing (Section \ref{sec:dt}). Before presenting a potentially buggy-exposing test
case to the developer, we apply a simple test case reduction algorithm to reduce the test case (Section \ref{sec:reduction}). Finally, to
minimize developer involvement, we use an incrementally built knowledge base to automatically analyze the testing outcomes to filter out
test cases that may trigger identical miscompilation behaviors seen before (Section \ref{sec:filtering}).

\begin{figure}[!t]
    \centering
    \subfigure[The extended AST]{
        \includegraphics[width=0.475\textwidth]{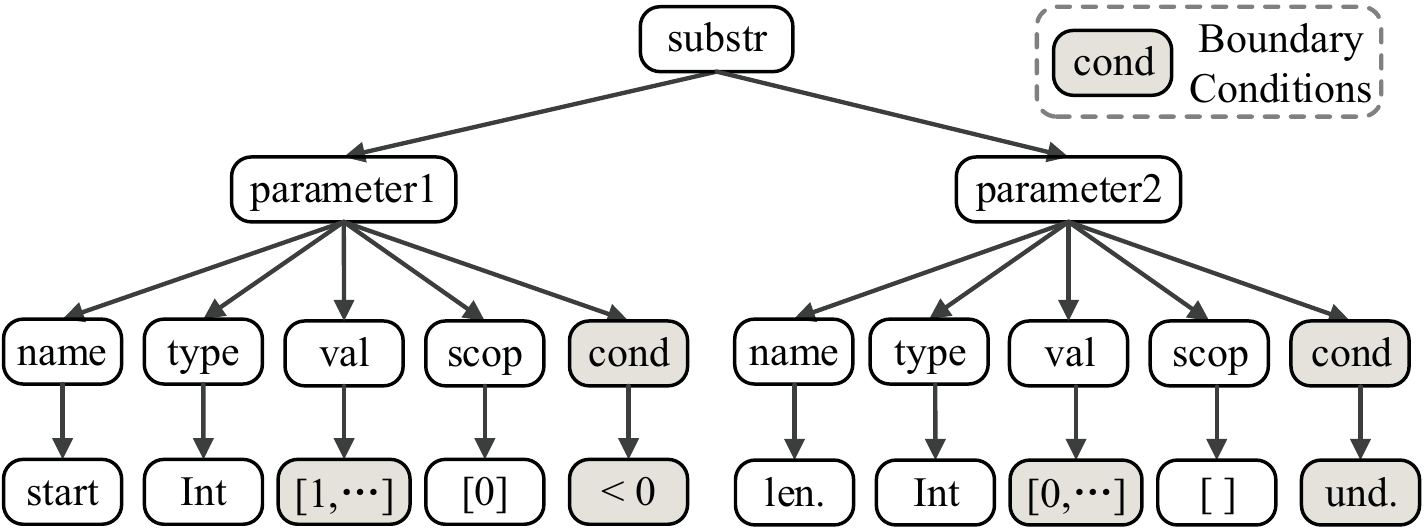}
        \label{fig:ast-example}
    }
    \vspace{-5mm}
    \subfigure[The JSON file of the AST in (a)]{
        \begin{minipage}[t]{0.5\textwidth}
\begin{lstlisting}[language=json,basicstyle=\scriptsize\ttfamily]
{
  "String.prototype.substr": [{
    "name": "start",
    "type": "integer",
    "values": [1, -1, "NaN", 0, "Infinity", "-Infinity"],
    "scopes": [0],
    "conditions": ["start < 0"]
  }, {
    "name": "length",
    "type": "integer",
    "values": ["undefined", "NaN", 0, "Infinity", "-Infinity"],
    "scopes": [],
    "conditions": ["length === undefined"]
  }]
}
\end{lstlisting}
            \label{fig:ast-storage}
        \end{minipage}
    }
    \caption{The AST (a) and its JSON format (b) for encoding specifications of the \texttt{substr} function in Figure~\ref{fig:pseudocode}.}
    \label{fig:ast}
    \vspace{-3mm}
\end{figure}

\subsection{Extracting \ecma Specification\label{sec:rulex}}
The goal of our \ecma parser is to extract specification rules from \ecma. In this work, we use the HTML version of the \ecma document.
Given the \ecma{} document, our parser first detects the scope of a function, class or object by analyzing the metadata. Specifically, we
use Tika~\cite{tika}, a content analysis library, with the help of hand-written regular expressions (regex) to extract the metadata (i.e.,
\ecma specification rules) of the HTML document.  For example, we use the regex \^{}\texttt{Let \${Var} be \${Func}\${Edn}\$} to extract
the initialization condition defined at line 4 of Figure~\ref{fig:pseudocode}. Our initial regular expression rules that account for ~80\%
of the rules used in experiments were written by a postgraduate student within a week. We have since improved and grown our regular
expression set.

The extracted specification is organized in the form of an abstract syntax tree (AST) where the root of a tree is an \ecma function, class,
or object and the children node is a rule defined for a specific object or function. We tag boundary conditions with special attributes on
the AST.  Figure \ref{fig:ast-example} shows the extended AST for the specification given in Figure~\ref{fig:pseudocode} where the boundary
conditions and literals are tagged with unique attributes. We then store the AST in the JSON format as illustrated in
Figure~\ref{fig:ast-storage} where the boundary conditions and value ranges are recorded. Since we primarily target \JS APIs, the extracted
rules are used to generate boundary conditions for mutating test programs related to \JS APIs.

Note that there are other \JS language specifications and definitions described in the natural language form. We do not extract rules from
such definitions. Overall, our extracted rules cover around 82\% of API and object specification rules in \ecma 10th Edition (2019
version).

\subsection{Test Program Generation\label{sec:generation}}

\cparagraph{Language model.} We employ a DNN-based language model to generate \JS test programs. Our \JS code generator is built by
fine-training a pre-trained GPT-2~\cite{gpt2model} (that was trained on natural language documents by independent researchers) on \JS
programs collected from open-source projects hosted on GitHub. The model requires each training input (i.e., \JS programs) to be
represented as a sequence of numerical values. To do so, we map every instruction, constant, and variable to an integer by looking up the
token in a vocabulary table. We construct the vocabulary using the  Byte Pair Encoding (BPE) tokenization method
~\cite{sennrich2016neural}, which is also used by GPT-2. This scheme works by first counting each word's frequency in the training data. It
then breaks each word into chunks (or subwords) based on the word frequency. For example, commonly seen language keywords, e.g.,
\texttt{var}, \texttt{for}, and \texttt{if}, will be tokenized as whole words, while rarer words like variable names will be broken into
smaller chunks (e.g., a few characters) and can be used to create the rest of the words. The algorithm tries to find a way to represent the
training dataset with the least amount of tokens and map each sub-word or token to an integer to be stored in the vocabulary table. This
scheme allows us to deal with a potential infinite number of words seen in real-life \JS programs by reusing tokens from a finite list of
subwords.

\cparagraph{Model training.} To collect training data to port the pre-trained GPT-2 for generating \JS programs, we developed an automated
tool to collect 140,000 \JS programs from 4,000 top-ranked (i.e., projects with the greatest numbers of stars) GitHub projects with \JS as
the main programming language. We use the collected \JS programs to update the weights of the last two fully-connected layers of the
pre-trained GPT-2 model  while keeping weights of other layers unchanged. The network is trained using the Adam
optimizer~\cite{kingma2015adam} for 100 epochs (over 150,000 iterations) with an initial learning rate of 0.0001 and decaying by 10\% every
epoch. Training the GPT-2 model took around 30 hours using four NVIDIA GTX 2080Ti desktop GPUs, which was a one-off cost. Note that we
provided the model with no prior knowledge of the structure or syntax of \JS.

\cparagraph{\JS program generation.} We use the trained GPT-2 to generate \JS test programs. Each of the test programs contains a \JS
function (e.g., function \texttt{foo} at lines 1--4 in Figure~\ref{fig:motivation_example}), which may invoke some standard \JS APIs. To
generate a test program, we feed the network with a randomly chosen seed generation header (e.g., ``\texttt{var a = function(assert)
\{}''). The seed generation header is chosen from a corpus of 2,000 function header samples, which were automatically collected from our
\JS training dataset. We ask the network to produce the next token based on the current text string, $s_c$. To choose the next token, we
employ a top-k sampling scheme, by randomly choosing a token from the top-k tokens that are predicted to have the highest possibilities by
following $s_c$ (we empirically set $k$ to $10$). We then append the chosen character to the existing string, $s_c$ and feed the new string
to the network to repeat the process to generate the next token. This generation process terminates when the brackets (i.e., `\{' and `\}'
) are matched, or a dedicated termination symbol ``<EOF>'' is produced by the language model, or the number of words of the synthetic
program is over 5,000. For each generated test program, we use JSHint~\cite{jshint} to \emph{statically} check its syntax to remove test
programs with syntax errors. Our current implementation also randomly keeps 20\% of the syntactically-invalid test programs for test runs.
A better approach for choosing syntax-incorrect programs for testing would consider program characteristics like API coverage and code
length. We leave this as our future work. Later in the paper, we show that compared to DeepSmith~\cite{cummins2018compiler}, GPT-2 can
model longer dependences among tokens in the source code, which in turns leads to a high success rate in producing syntactically correct
test programs (see Section~\ref{sec:tcg}).

\subsection{\ecma{}-guided Test Data Generation\label{sec:tdg}}
Our test data is embedded into the \JS code by assigning values to variables that are passed to a \JS function. In addition to generating
variables, we also generate code to call functions with supplied parameters and print out the results. For example, lines 5--9  in Figure
\ref{fig:motivation_example} show the test data and the corresponding test-driven code produced by \SystemName.

Algorithm~\ref{alg:ecma} presents our test data generation. Our approach takes in a \JS test case program and outputs multiple test
cases.  For each statement in the input \JS test program, we first check if the statement contains a function invocation, and then locate the \JS API definitions and the potential arguments of the API, by using the API
name to look up our ECMA-262 database (lines 1--7). As the \emph{training corpus} for our test program generation model
contains human-written programs with code and data of various API calling patterns, the test programs generated by \SystemName
may already contain code and data to invoke a \JS API in different ways. Our approach keeps these test cases but will use the \ecma rules to
mutate the values assigned to function arguments to generate additional test data samples. Specifically, it uses the ECMA-262 rules extracted offline to determine how many parameters
should be passed to a function, and the type and value for each parameter. For each parameter type, we mutate the values based on (1)
boundary conditions according to the \ecma{} specification (e.g., in Figure~\ref{fig:pseudocode}, arguments \texttt{len} is set to \texttt{undefined}), and (2) normal conditions (using random values). To mutate the variable values, we
associate an argument passed to a function with its definition by traversing the \JS program's control and data flow graph (line 8). These parameter values and the input \JS test program are then store in our list of test case programs (line 9).

\begin{algorithm}[!t]
    \centering
    \caption{\ecma{}-guided Test Data Generation}
    \label{alg:ecma}
    \small
    \begin{algorithmic}[1]
       \REQUIRE~~\\
            $t_{prog}$: A \JS test case program\\
            $ecma$: A supported edition of \ecma{}\\
        \ENSURE~~\\
            $T_{new}$: A list of mutated test case programs \\
        \STATE $Specs \leftarrow$ extractFuncSpecs($ecma$)\\
        \STATE Let $T_{new}$ be a list\\
           \FOR{$st\in$ $t_{prog}$}
               \IF{isFunc($st$)}
                \STATE $funcName \leftarrow$ getFuncName($st$)\\
           \IF{$Specs$.containsName($funcName$)}
                 \STATE $spec_{func} \leftarrow$  getSpecs($Specs$, $st$)\\
                  \STATE $t_{new} \leftarrow$  mutate($t_{prog}$, $func$, $spec_{func}$)\\
                   \STATE $T_{new}$.append($t_{new}$)\\
                \ENDIF
               \ENDIF
            \ENDFOR
        \RETURN $T_{new}$
    \end{algorithmic}
\end{algorithm}

\subsection{Differential Testing\label{sec:dt}}
We employ the established differential testing methodology \cite{chen2020survey} to expose \JS compiler defects by running a test case
across multiple \JS engines (or Testbeds). We use a majority voting scheme to determine which compiler's behavior deviates from others by
comparing the results of compilation and execution. Differential testing typically requires the test program to yield a deterministic,
well-defined  outcome \cite{clsmith}. The use of ECMA-262 specifications to generate test data enable us to create test cases with expected
deterministic behaviors. The test programs generated by our language model may have non-deterministic outcomes like floating-point rounding
errors. However, we did not experience this problem in our test runs - if the behavior of a \JS engine deviates from others, it was
typically due to a bug of the compiler implementation during our test runs.

\begin{figure}
    \centering
    \includegraphics[width=0.45\textwidth]{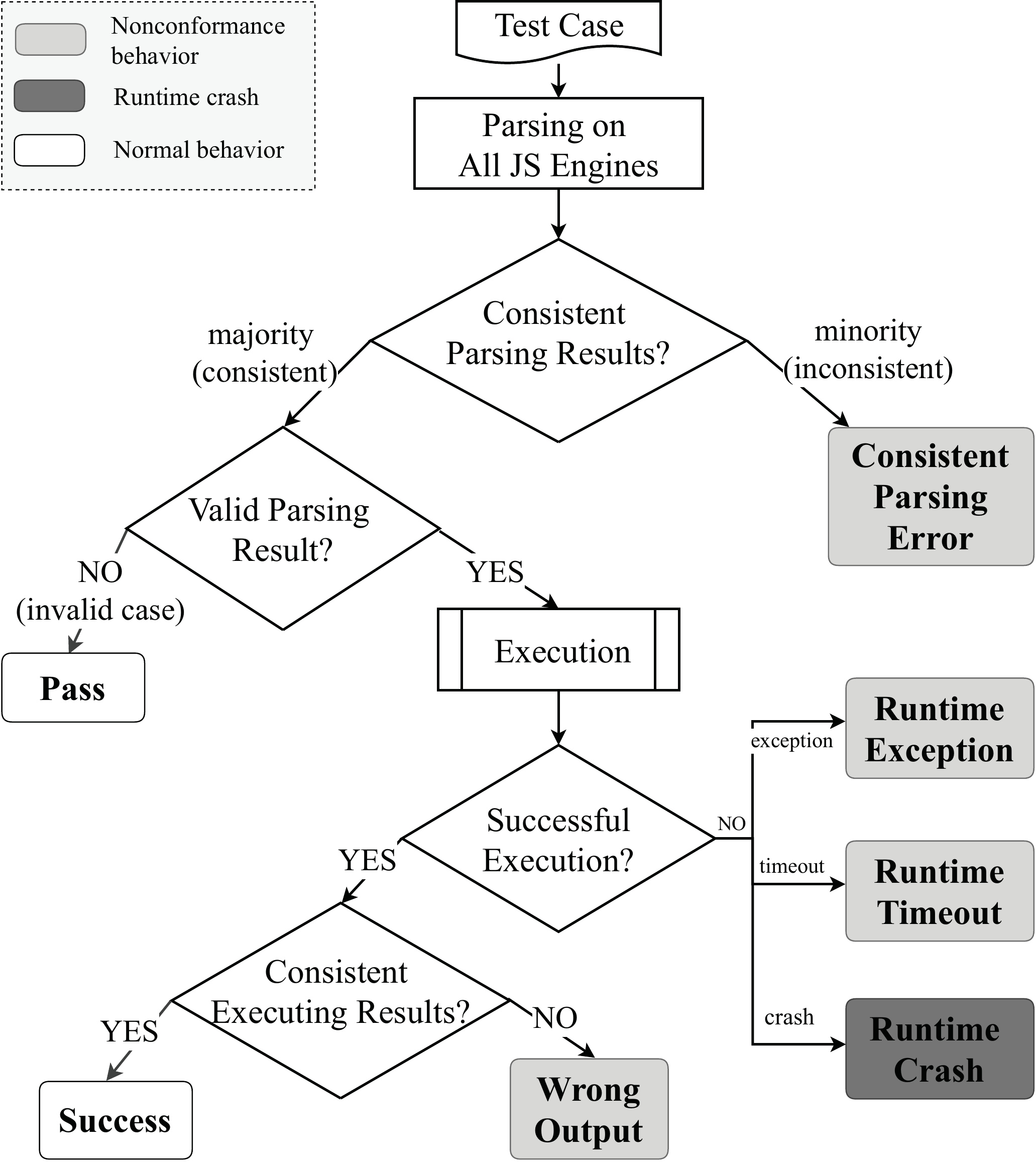}
    \vspace{-2mm}
    \caption{Possible outcomes when executing a test case.}
    \label{fig:differential}
    \vspace{-3mm}
\end{figure}

Executing a test case on a \JS engine leads to one of seven possible outcomes, illustrated in Figure~\ref{fig:differential}. A consistent
parsing error occurs when the parsing results (both successful or not successful) are inconsistent. We ignore test \JS programs that fail
to be successfully parsed by all test engines. The successfully parsed \JS programs will be executed, which can lead to five runtime
outcomes. A wrong output occurs if the execution results are inconsistent among \JS engines when a deterministic behavior is expected
according to ECMA-262.  This often happens when a \JS engine produces a result (e.g., throwing an exception when it should not) that is
different from the expected behavior defined in ECMA-262. A runtime crash occurs if the \JS engine crashes when executing a test case. A
runtime timeout happens if a \JS test case fails to terminate when it is running with a \JS engine within a period of $2t$, where $t$ is
the longest time for all other \JS engines to return the result. Note that we ignore test cases where all \JS engines do not terminate
within ten minutes because this is likely to be due to a large or infinite loop in the test program. Finally, we consider a test case to be
a passing one if all tested \JS engines can successfully execute it, and the executions lead to a consistent outcome.

When evaluating the outcomes of test cases, compile crash and timeout outcomes are of immediate interest, indicative of erroneous compiler
behavior. For all other outcomes, we use differential testing methodology to confirm anomalous behavior. Specifically, we compare the
results obtained for a test case using multiple \JS engines, including \JS engines in the same family but from different trunk builds. We
look for test cases where a \JS compiler's behavior deviates from all others, which can then be investigated by developers.

\vspace{-2mm}
\subsection{Test Case Reduction\label{sec:reduction}}
To help developers in examining bug-exposing test cases, we develop a simple yet effective approach to reduce the size of a test case that
triggers an anomalous compiler behavior. Our idea is to traverse the abstract syntax tree of the input test program to iteratively remove
code structures and test if the resulted program can still trigger the same compilation or execution outcome. We repeat this process until
a fixpoint where no additional reduction from can be done while still reproducing the same anomalous behavior as the original test case
does. We note that there are other test case reduction tools like HDD~\cite{misherghi2006hdd} and Perses~\cite{sun2018perses} to be used
for the same purpose.

\begin{figure}
    \centering
    \includegraphics[width=0.45\textwidth]{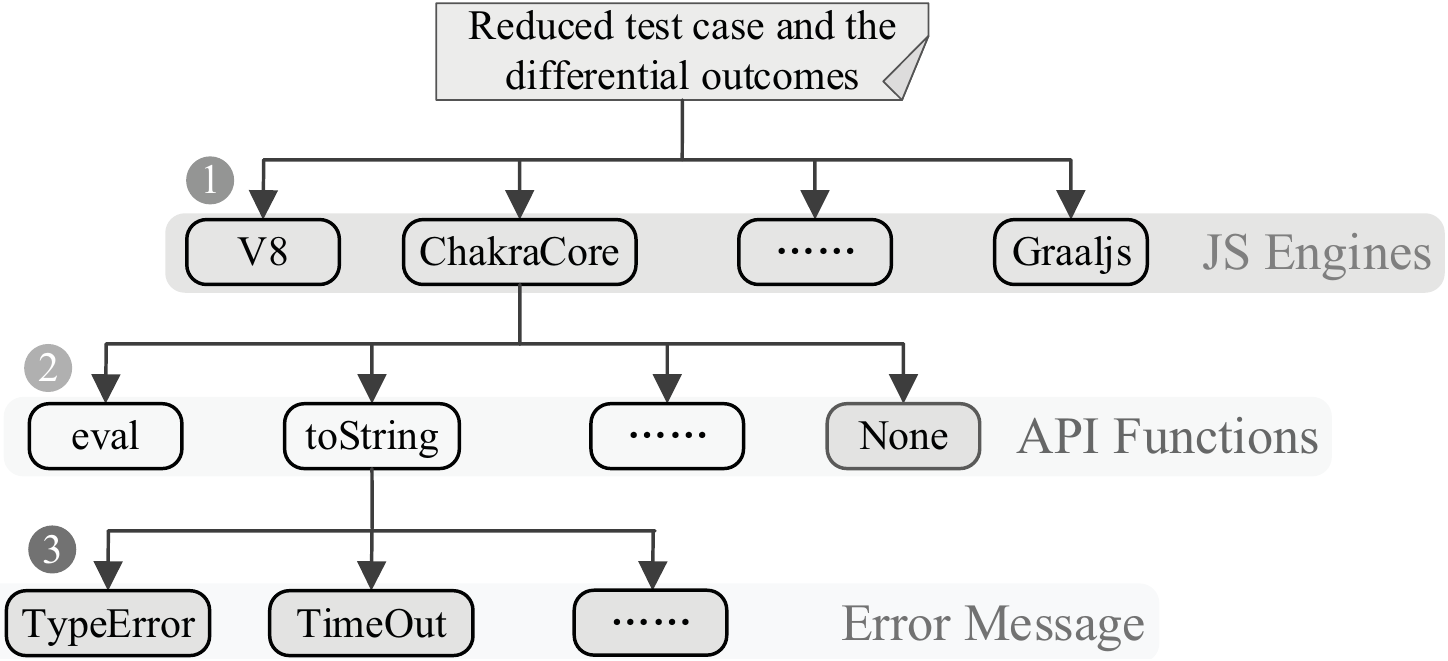}
    \caption{Our tree-based identical bug filter.}
    \vspace{-2mm}
    \label{fig:filter}
    \vspace{-5mm}
\end{figure}

\subsection{Filtering Identical Miscompilation\label{sec:filtering}}
Random program generation often produces test cases that trigger the same compiler bugs. To reduce developer involvement, we employ an
automated scheme to filter out test cases that are likely to cause identical miscompilation behaviors. To do so, we construct a simple
tree-based knowledge base from historical bug-exposing test cases. Our tree-based classifier to identify identical bugs is inspired by
prior work on employing decision trees for predictive modeling~\cite{grewe2013portable,wang2014automatic,wang2018machine}. We choose this
technique because the model is interpretable.

Figure~\ref{fig:filter} presents a high-level overview of our tree-based knowledge base, which provides three layers to predict if a test
case triggers a bug that was seen by an already analyzed test case. Every decision node in the top layer corresponds to a \JS engine, which
checks if the test case triggers the same bug seen for a \JS engine. Likewise, the API function nodes in the second layer check if the test
case triggers a bug of a specific \JS API. If the test case does not contain a \JS API function, it will be classified into the
\texttt{None} leaf node in the second layer. The leaf nodes in the last layer group the differential results based on the miscompilation
behaviors (such as TypeError, TimeOut, Crash, etc.) that the tested \JS engine yields. If a test case triggers a
buggy behaviour during test runs as described in Section \ref{sec:dt}, we then traverse this tree to see if there already exists a path
that gives identical information as exposed by the test case. If so, we consider a previously identified bug is found.  If not, we consider
a new bug is triggered for a given \JS engine and \JS API. For the latter case, we then add a new leaf node to the tree according to the
bug the test case triggers.

We have so far created a knowledge base consisting of 2,300 leaf decision nodes. In our current implementation, the tree is realized as a
set of rules written in Python. These rules already helped us filter out over ten of thousands of repeated miscompilation behaviors,
significantly reducing the overhead and human involvement of examining test cases to confirm bugs.

\vspace{-2mm}
\section{Experimental Setup}
\begin{table}[!t]
    \centering
    \caption{\JS engines we have tested.}
    \label{tab:compilers}
    \vspace{-2mm}
    \scriptsize
    \begin{tabularx}{0.48\textwidth}{XXXrr}
    \toprule
    \textbf{\JS Engine} & \textbf{Versions} & \textbf{Build No.} & \textbf{Release Date} & \textbf{Supported ES Spec.} \\
    \midrule
    \rowcolor{Gray} & V8.5 & d891c59 & Jun. 2020 &  \\
    \rowcolor{Gray} & V8.5 & e39c701 & Aug. 2019 & \\
    \rowcolor{Gray} \multirow{-3}{*}{\rotatebox[origin=c]{90}{V8}} & V8.5 & 0e44fef & Apr. 2019 & \multirow{-3}{*}{ES2019} \\

    \multirow{5}{*}{\rotatebox[origin=c]{90}{ChakraCore}} & v1.11.19 & 5ed2985 & May 2020 &  \\
    & v1.11.16 & eaaf7ac & Nov. 2019 &  \\
    & v1.11.13 & 8fcb0f1 & Aug. 2019 &  \\
    & v1.11.12 & e1f5b03 & Aug. 2019 &  \\
    & v1.11.8 & dbfb5bd & Apr. 2019 & \multirow{-5}{*}{ES2019} \\

    \rowcolor{Gray} & 261782 & dbae081 & May 2020 &  \\
    \rowcolor{Gray} & 251631 & b96bf75 & Oct. 2019 &  \\
    \rowcolor{Gray} & 246135 & d940b47 & Jun. 2019 &  \\
    \rowcolor{Gray} \multirow{-4}{*}{\rotatebox[origin=c]{90}{JSC}}& 244445 & b3fa4c5 & Apr. 2019 & \multirow{-4}{*}{ES2019} \\

    \multirow{7}{*}{\rotatebox[origin=c]{90}{SpiderMonkey}} & v78.0 & C69.0a1 & Jun. 2020 &  \\
     & gecko-dev & 2c619e2 & May 2020 &  \\
    & gecko-dev & 201255a & Jun. 2019 &  \\
    & v60.1.1 & mozjs60.1.1pre3 & Jul. 2018 &  \\
    & v52.9 & mozjs52.9.1pre1 & Jul. 2018 &  \\
    & v38.3.0 & mozjs38.3.0 & Oct. 2017 &  \\
    & v1.7.0 & js-1.7.0 & Sep. 2017 & \multirow{-7}{*}{ES2018/2019} \\

   \rowcolor{Gray} & v1.7.12 & d4021ee & Jan. 2020 &  \\
   \rowcolor{Gray} & v1.7.11 & f0e1c63 & May 2019 &	 \\
   \rowcolor{Gray} & v1.7.10 & 1692f5f & May 2019 &  \\
   \rowcolor{Gray} & v1.7.9	& 3ee580e & Mar. 2018 &  \\
   \rowcolor{Gray} & v1.7R5 & 584e7ec & Jan. 2015 &  \\
   \rowcolor{Gray} & v1.7R4 & 82ffb8f & Jun. 2012 &  \\
   \rowcolor{Gray} \multirow{-7}{*}{\rotatebox[origin=c]{90}{Rhino}} & v1.7R3 &	d1a8338 & Apr. 2011 & \multirow{-7}{*}{ES2015} \\

   \multirow{5}{*}{\rotatebox[origin=c]{90}{Nashorn}} & v13.0.1 & JDK13.0.1 & Sep. 2019 &	 \\
    & v12.0.1 & JDK12.0.1 & Apr. 2019 &  \\
    & v11.0.3 & JDK11.0.3 & Mar. 2019 &  \\
    & v1.8.0\_201 & JDK8u201 & Jan. 2019 &  \\
    & v1.7.6 & JDK7u65 & May 2014 & \multirow{-5}{*}{ES2011/2015} \\

    \rowcolor{Gray} & v0.6.0 & b6530ae & May 2020 &  \\
    \rowcolor{Gray} & v0.4.0 & 044cf4b & Dec. 2019 &  \\
    \rowcolor{Gray} & v0.3.0 & 3826084 & Sep. 2019 &  \\
    \rowcolor{Gray} \multirow{-4}{*}{\rotatebox[origin=c]{90}{Hermes}} & v0.1.1 & 3ed8340 & Jul. 2019 & \multirow{-4}{*}{ES2015} \\

   \multirow{9}{*}{\rotatebox[origin=c]{90}{JerryScript}} & v2.3.0 & bd1c4df & May 2020 &  \\
   & v2.2.0 & 996bf76 & Nov. 2019 &  \\
   & v2.2.0 & 7df87b7 & Oct. 2019 &  \\
   & v2.1.0 & 84a56ef & Oct. 2019 &  \\
   & v2.1.0 & 9ab4872 & Sep. 2019 &  \\
   & v2.0 & 351acdf & Jun. 2019 &  \\
   & v2.0 & b6fc4e1 & May 2019 &  \\
   & v2.0 & 40f7b1c & Apr. 2019 &  \\
   & v2.0 & e944cda & Apr. 2019 & \multirow{-9}{*}{ES2011/2015} \\

    \rowcolor{Gray} & 2020-04-12 & 1722758 & Apr. 2020 &  \\
    \rowcolor{Gray} & 2020-01-05 & 91459fb & Jan. 2020 &  \\
    \rowcolor{Gray} & 2019-10-27 & eb34626 & Oct. 2019 &  \\
    \rowcolor{Gray} & 2019-09-18 & 6e76fd9 & Sep. 2019 &  \\
    \rowcolor{Gray} & 2019-09-01 & 3608b16 & Sep. 2019 &  \\
    \rowcolor{Gray} \multirow{-6}{*}{\rotatebox[origin=c]{90}{QuickJS}} & 2019-07-09 & 9ccefbf & Jul. 2019 & \multirow{-6}{*}{ES2019} \\

    Graaljs & v20.1.0 & 299f61f & May 2020 & ES2020 \\
    \bottomrule
    \end{tabularx}
    \vspace{-5mm}
\end{table}

\subsection{\JS Engines}
Table \ref {tab:compilers} lists the \JS engines and versions used in our evaluation. We apply \SystemName to ten \JS engines and test
several trunk branches of each engine. All the tested \JS engines were claimed to be compatible with a version of ECMA-262. In evaluation,
we ensure that we only test each \JS engine against the corresponding
supported ECMA-262 edition.
In total, we have tested 51 \JS engine-version configurations.

\subsection{Testbeds}
For each \JS configuration, we create two testbeds. In the first, the engine runs under the normal mode. In the second, the engine runs
under the strict mode. The ``strict mode'' is introduced since ECMAScript 5, which allows one to run a \JS program in a ``strict''
operating context. Programs running in the strict mode can have different semantics from normal code. For example, the strict mode
eliminates some \JS silent errors by forcing them to throw errors per the ECMA-262 standard. This testing mechanism gives us a total of 102
testbeds (51 \JS engine version configurations $\times$ 2 testbeds per configuration) to evaluate. For the remainder of the paper, unless
state otherwise, the bugs are reported to be found under both the normal and the strict modes.

\vspace{-2mm}
\subsection{Test Case Generation}
We use \SystemName to produce a total of 300k synthetic test cases. We use JSHint~\cite{jshint}, a static \JS parser to remove syntactically incorrect test
programs. However, we still keep a small number (10k) of randomly chosen test cases  that are considered to have syntax errors to test the
parser of a \JS engine. On average, we keep 250k test cases.

\vspace{-2mm}
\subsection{Competitive Baselines}
We compare \SystemName against both generation- and mutation-based fuzzers. Specifically, we compare \SystemName with DeepSmith
\cite{cummins2018compiler}, a closely related DNN-based program generator for compiler fuzzing. We also compare \SystemName to four
mutation-based \JS fuzzers: Fuzzilli \cite{gross2018fuzzil}, CodeAlchemist~\cite{han2019codealchemist}, DIE~\cite{park2020fuzzing} and
Montage~\cite{son2020montage}, where the last three represent the state-of-the-art \JS compiler fuzzers.

\vspace{-1mm}
\subsection{Hardware Platforms}
Our evaluation platform is a multi-core server with a 3.6GHz 8-core (16 threads) Intel Core i7 CPU, four NVIDIA GTX 2080Ti GPUs and 64GB of
RAM, running Ubuntu 18.04 operating system with Linux kernel 4.15. All DNN models run on the native
hardware using the GPUs. For fuzzing tests, we run each testbed in a Docker container (version 19.03.5) so that we can run 16 fuzzing processes
simultaneously.

\begin{table}[!t]
    \centering
    \caption{Bug statistics for each tested \JS engine.}
    \scriptsize
    \vspace{-2mm}
    \label{tab:statistics}
    \begin{tabularx}{0.48\textwidth}{Xrrrr}
        \toprule
        & & \multicolumn{2}{c}{\textbf{\#Confirmed}} & \\
        \cline{3-4}\multirow{-2}{*}{\textbf{\JS Engine}} & \multirow{-2}{*}{\textbf{\#Submitted}} & \textbf{\#Verified} & \textbf{\#Fixed in \#Verf.} & \multirow{-2}{*}{\textbf{\#Acc. by Test262}}\\
        \midrule
        \rowcolor{Gray} V8 & 4 & 4 & 3 & 1 \\
        ChakraCore & 7 & 7 & 5 & 1 \\
        \rowcolor{Gray} JSC & 12 & 11 & 11 & 3 \\
        SpiderMonkey & 3 & 3 & 3 & 0 \\
        \rowcolor{Gray}Rhino & 44 & 29 & 29 & 4 \\
        Nashorn & 18 & 12 & 2 & 1 \\
       \rowcolor{Gray} Hermes & 16 & 16 & 15 & 4 \\
        JerryScript & 35 & 31 & 31 & 3 \\
        \rowcolor{Gray}QuickJS & 17 & 14 & 14 & 4 \\
        Graaljs & 2 & 2 & 2 & 0 \\
      \rowcolor{Gray}  \textbf{Total} & \textbf{\SBugs} & \textbf{\VBugs} & \textbf{\FBugs} & \textbf{\TSC} \\
        \bottomrule
    \end{tabularx}
    \vspace{-2mm}
\end{table}

\section{Experimental Results}
From May 2019, we have started experimenting with and refining \SystemName to find bugs in  Rhino and then gradually extended our tests to
other \JS engines. From May 2020, we started our extensive testing of all \JS engines. In total, we test each \JS testbed through 200 hours
automated runs on 250k automatically generated test cases. Unless stated otherwise, \SystemName code listings are presented verbatim, with
only minor formatting changes applied to save space. No manual test case reduction was performed.

\begin{table}[!t]
    \centering
    \caption{The number of bugs found per \JS engine version.}
    \label{tab:versions}
    \scriptsize
    \begin{tabularx}{0.48\textwidth}{XXrrrr}
        \toprule
        & & & \multicolumn{3}{c}{\textbf{\#Confirmed}} \\
        \cline{4-6}\multirow{-2}{*}{\textbf{JS Engine}} & \multirow{-2}{*}{\textbf{Versions}} & \multirow{-2}{*}{\textbf{\#Submitted}} & \textbf{\#Verified} & \textbf{\#Fixed} & \textbf{\#New} \\
        \midrule
        \rowcolor{Gray} v8 & V8.5 & 4 & 4 & 3 & 4 \\

        \multirow{4}{*}{ChakraCore} & v1.11.16 & 3 & 3 & 1 & 3 \\
        & v1.11.13 & 1 & 1 & 1 & 0 \\
        & v1.11.12 & 1 & 1 & 1 & 1\\
        & v1.11.8 & 2 & 2 & 2 & 2\\

        \rowcolor{Gray} & 261782 & 1 & 1 & 1 & 1\\
        \rowcolor{Gray} & 251631 &2 & 1 & 1 & 1\\
        \rowcolor{Gray} & 246135 & 8 & 8 & 8 & 6\\
        \rowcolor{Gray} \multirow{-4}{*}{JSC}& 244445 & 1 & 1 & 1 & 0\\

        \multirow{3}{*}{SpiderMonkey} & v52.9 & 1 & 1 & 1 & 0 \\
        & v38.3 & 1 & 1 & 1 & 0\\
        & v1.7 & 1 & 1 & 1 & 0\\

      \rowcolor{Gray} & v1.7.12 & 25 & 19 & 19 & 19 \\
      \rowcolor{Gray} & v1.7.11 & 17 & 8 &	8 & 4\\
      \rowcolor{Gray} \multirow{-3}{*}{Rhino} & v1.7.10 & 2 & 2 & 2 & 2\\

      \multirow{2}{*}{Nashorn} & v13.0.1 & 4 & 4 & 0 & 4 \\
        & v12.0.1 & 14 & 8 & 2 & 7 \\

        \rowcolor{Gray} & v0.6.0 & 2 & 2 & 2 & 2\\
        \rowcolor{Gray} & v0.4.0 & 1 & 1 & 0 & 1\\
        \rowcolor{Gray} & v0.3.0 & 6 & 6 & 6 & 5\\
        \rowcolor{Gray} \multirow{-4}{*}{Hermes} & v0.1.1 & 7 & 7 & 7 & 4\\

      \multirow{5}{*}{JerryScript} & v2.3.0 & 2 & 2 & 2 & 2 \\
      & v2.2.0 & 18 & 16 & 16 & 15 \\
      & v2.1.0 & 6 & 5 & 5 & 4 \\
      & v2.0 & 8 & 7 & 7 & 7 \\
      & v1.0 & 1 & 1 & 1 & 1 \\

        \rowcolor{Gray} & 2020-04-12 & 1 & 1 & 1 & 1 \\
        \rowcolor{Gray} & 2020-01-05 & 2 & 2 & 2 & 2\\
        \rowcolor{Gray} & 2019-10-27 & 4 & 3 & 3 & 3\\
        \rowcolor{Gray} & 2019-09-18 & 3 & 1 & 1 & 1\\
        \rowcolor{Gray} & 2019-09-01 & 4 & 4 & 4 & 4\\
        \rowcolor{Gray} \multirow{-6}{*}{QuickJS} & 2019-07-09 & 3 & 3 & 3 & 1\\

        Graaljs & v20.1.0 & 2 & 2 & 2 & 2 \\
        \midrule
        \textbf{Total} & \textbf{33} & \textbf{158} & \textbf{129} & \textbf{115} & \textbf{109} \\
        \bottomrule
    \end{tabularx}
    \vspace{-4mm}
\end{table}

\cparagraph{Highlights.} As of November 2020, we have indentified\footnote{A list of our bug reports can be found at:
\url{https://github.com/NWU-NISL-Fuzzing/COMFORT/blob/main/artifact_evaluation/docs/Bug-List.md}} \SBugs unique bugs (of which 109 were
found to be newly discovered bugs by developers). So far, \VBugs have been verified, of which \FBugs have been fixed by the developers. For
the remaining 29 unverified bugs, 9 were rejected by the developers as the feature was either not clearly defined in ECMA-262 or not
supported by the compiler version; and others were either under discussion or yet to be confirmed. Many of the bugs discovered by
\SystemName were not exposed by the state-of-the-art compiler fuzzing methods. Moreover, 21 of the \SystemName-generated test cases have
been added to Test262, of which we submitted 18 test cases, and the JSC developers submitted 3 \SystemName-generated test cases after we
have reported the bug.

\subsection{Quantitative Results}
This subsection presents various summary statistics on results from our \JS compiler testing effort.

\subsubsection{Bug count} Table~\ref{tab:statistics} gives the distribution of the \SystemName-exposing bugs across the tested \JS engines.
Each of all the ten evaluated \JS engines has at least one conformance bug even though all of them claimed to adhere to the \ecma{}
specification tested. This shows the prevalence of conformance bugs across different \JS engines. Note that when a bug is confirmed and
triaged, it corresponds to a new defect. Therefore, all confirmed bugs that we reported were unique and independent. Although we have
checked that all reported bugs have different symptoms, some of them were actually linked to the same root cause. A total of eight such bug
reports were later marked as \emph{duplicate} by developers.

Note that not all confirmed bugs of Nashorn were fixed because their developers ceased maintaining this engine after June 2020. It is not
surprising that \JS engines like V8 and SpiderMonkey with a long development and testing history and a larger developer community have
fewer bugs than some newer open-source \JS engines like JerryScript. We found the SpiderMonkey implementation to be well-conformed to
\ecma, and we only found three conformance bugs in a previous release.

Table~\ref{tab:versions} shows the number of unique bugs found per \JS engine version. Note that some of the bugs may exist across
different versions of the same \JS engine. For clarity, we only attribute the discovered bugs to the earliest bug-exposing version used in
our evaluation. In total, \SystemName discovered 38 new bugs from the latest versions listed in Table~\ref{tab:versions}.  Furthermore,
\SystemName has also found a considerable number of bugs in stable releases that had been latent for years.

It is worth mentioning that for Rhino version 1.7.12 and JerryScipt version 2.2.0, \SystemName found over 15 conformance bugs, far more than the number of bugs
found in other versions of these two engines. This is because Rhino and JerryScipt initially supported \ecma version 5 but have recently
added the support for \ecma version 6 at these two versions. This larger number of conformance bugs introduced when switching to a new
\ecma edition is expected as some of the API implementations in \ecma version 6 are different from those in version 5.

\begin{table}[!t]
    \centering
    \caption{Bug statistics for each group.}
    \label{tab:bug-type}
     \vspace{-3mm}
    \scriptsize
    \begin{tabular}{m{1.5cm}rrrr}
        \toprule
         \textbf{Category}& \textbf{\#Submitted} & \textbf{\#Confirmed} & \textbf{\#Fixed} & \textbf{\#Acc. by Test262} \\
        \hline
        \rowcolor{Gray} Test program generation & 97 & 78 & 67 & 5 \\
        \ecma{} guided mutation & 61 & 51 & 48 & 16 \\
        \bottomrule
    \end{tabular}
    \vspace{-3mm}
\end{table}

\subsubsection{Bug types} We distinguish two kinds of bugs: (1) ones that manifest through our test program generation (Section \ref
{sec:generation}) and (2) ones that manifest by exploiting the \ecma specification to generate test data (Section \ref{sec:tdg}).
Table~\ref{tab:bug-type} classifies the bugs found by \SystemName into these two groups.
Our GPT-2 test program generator is effective in generating bug-exposing test cases which contribute to 97 of the submitted bugs.  By
exploiting \ecma, \SystemName is able to discover further 61 bugs of which 51 have been confirmed. All of the bugs discovered in this
category are standard conformance bugs. Furthermore, 16 of the automatically generated test cases under this category have been added into
Test262. This shows the usefulness and importance in exploiting the language standard for exposing compiler bugs.

\begin{table}[!t]
    \centering
    \caption{Statistics on top-10 buggy object types.}
    \label{tab:api-distribution}
     \vspace{-3mm}
    \scriptsize
    \begin{tabularx}{0.48\textwidth}{Xrrr}
        \toprule
        \textbf{API Type} & \textbf{\#Submitted} & \textbf{\#Confirmed} & \textbf{\#Fixed} \\
        \midrule
        \rowcolor{Gray} \texttt{Object} & 23 & 21 & 18 \\
        \texttt{String} & 22 & 20 & 19 \\
        \rowcolor{Gray} \texttt{Array} & 17 & 12 & 9 \\
        \texttt{TypedArray} & 8 & 5 & 5 \\
        \rowcolor{Gray} \texttt{Number} & 5 & 4 & 4 \\
        \texttt{eval function} & 4 & 4 & 4 \\
        \rowcolor{Gray} \texttt{DataView} & 4 & 2 & 2 \\
        \texttt{JSON} & 3 & 3 & 2 \\
        \rowcolor{Gray} \texttt{RegExp} & 2 & 2 & 1 \\
        \texttt{Date} & 2 & 1 & 1 \\
        \rowcolor{Gray} \textbf{Total} & \textbf{90} & \textbf{74} & \textbf{65} \\
        \bottomrule
    \end{tabularx}
    \vspace{-3mm}
\end{table}

\subsubsection{API distribution} Table~\ref{tab:api-distribution} groups the buggy \JS API implementations found by \SystemName according to
the object type.
To investigate the APIs that are more likely to contain conformance bugs, we analyze the top-10 buggy object types.
Most of the bugs found by \SystemName are
object and string operations. This is due to the large number of standard \JS APIs provided for these two data types. For example, eight of the
confirmed and fixed bugs found by \SystemName were found for \texttt{String.prototype.replace()} due to improper handling the type and
number of the arguments. We have also found four confirmed bugs for the \texttt{eval} function for ChakraCore, Hermes and QuickJS due to the
inappropriate implementations for expression parsing and evaluation. One of such examples is given in Listing~\ref{list:eval}. This table shows that
conformance bugs can be found on a range of APIs.

\subsubsection{Affected compiler components} 
\begin{figure}
    \centering
    \includegraphics[width=0.42\textwidth]{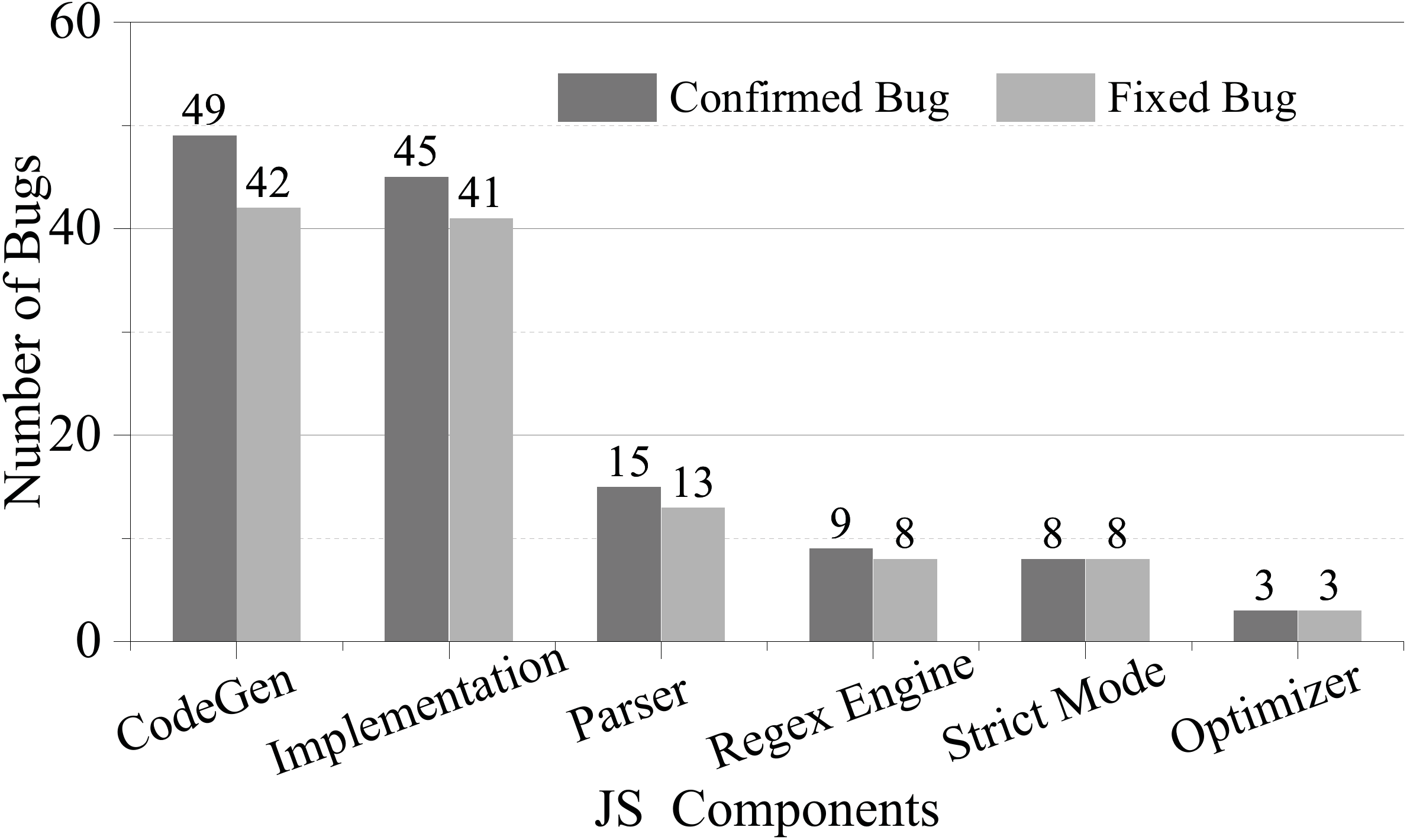}
    \vspace{-3mm}
    \caption{\#\SystemName-found bugs per compiler component.}
    \label{fig:components}
    \vspace{-5mm}
\end{figure}

We grouped the \SystemName-discovered bugs according to the typical \JS engine components: code
generation (CodeGen), API and library implementation (Implementation), Parser, the regular expression engine (Regex Engine), and Optimizer. We
also list bugs found solely in the strict mode. Figure~\ref{fig:components} shows the number of bugs discovered by \SystemName for each
component. Most of the bugs found were due to erroneous implementations in the back-end code generator. Bugs due to the library and API
implementation are also common - 45 confirmed and 41 fixed bugs belong to this category. According to the developer feedback, this is often
due to an oversight or misunderstanding of the ECMA-262 specification. 

\subsection{Bug Examples}
\SystemName is capable of finding diverse types of \JS engine bugs. To provide a glimpse of the diversity of the uncovered bugs, we
highlight here several of the \SystemName-produced test cases that expose a \JS compiler bug.

\cparagraph{\texttt{defineProperty} API.} All versions of V8 tested fail to correctly compile the test case shown in
Listing~\ref{list:defineProperity}. This test case contains a ``type error'' because the \texttt{length} property at line 3 is \emph{not}
configurable but the code tries to change it. Such an attempt should lead to a TypeError exception, but V8 allows the code to be
successfully compiled. This bug is exposed by generating a test data to manipulate a non-configurable property of the array object
according to ECMA-262. We were the first to report this bug. This test case also exposes a bug of Graaljs for the same reason.

\begin{lstlisting}[basicstyle=\footnotesize, frame=lines, caption={All versions of V8 tested fail to throw a TypeError exception for this
test case.},upquote=true,captionpos=b,label={list:defineProperity}]
var foo = function(){
  var arrobj = [0, 1];
  Object.defineProperty(arrobj, "length", {
    value: 1, configurable: true
  });
};
foo();
\end{lstlisting}

\cparagraph{Performance bug.} The test code in Listing~\ref{list:performance} exposes a performance issue of Hermes's memory allocation
policy. Hermes took more than half an hour to execute the code while other \JS engines use less than a second. Due to its memory allocation
strategy, Hermes has to relocate the array for every new element inserted on the left (when the array elements are added in reverse order),
leading to significant runtime overhead with a large array. This test case was generated by our GPT-2 test program generator. This bug
affects all versions prior to V0.3.0, and we were the first to report this. Our bug report was welcomed and quickly fixed by the Hermes
developers.

\begin{lstlisting}[basicstyle=\footnotesize, frame=lines, caption={Hermes took 30+ minutes to execute this code while other engines took less
than a second.},upquote=true,captionpos=b,label={list:performance}]
var foo = function(size) {
  var array=new Array(size);
  while (size--){
    array[size] = 0;
  }
};
var parameter = 904862;
foo(parameter);
\end{lstlisting}

\cparagraph{\texttt{Uint32Array}.} SpiderMonkey prior to v52.9 incorrectly threw a TypeError exception for the code shown in
Listing~\ref{list:unit32array}. This is because it does not convert the function argument to an integer type per ECMA-262
before calling the \texttt{Uint32Array} API at line 2. A standard conforming implementation would convert 3.14 to 3 to be used as the array
length.
\begin{lstlisting}[basicstyle=\footnotesize, frame=lines, caption={SpiderMonkey before v52.9
incorrectly throws a TypeError exception.},upquote=true,captionpos=b,label={list:unit32array}]
var foo = function(length){
  var array = new Uint32Array(length);
  print(array.length);
};
var parameter = 3.14;
foo(parameter);
\end{lstlisting}

\cparagraph{\texttt{Number.prototype.toFixed}.} Rhino compiled and executed the test case in Listing~\ref{list:toFixed} to produce an
 output of ``-634619'', when it should throw a RangeError exception. ECMA-262 states \texttt{Number.prototype.toFixed} only takes
a value between 0 and 20. Hence, this is a conformance bug.

\begin{lstlisting}[basicstyle=\footnotesize, frame=lines, caption={Rhino fails to throw a RangeError exception for this test
case.},upquote=true,captionpos=b, label={list:toFixed}]
var foo = function(num){
  var p = num.toFixed(-2);
  print(p);
};
var parameter = -634619;
foo(parameter);
\end{lstlisting}

\vspace{-2.5mm} \cparagraph{\texttt{\%TypedArray\%.prototype.set}.} The tested JSC trunk builds prior to v261782 threw a TypeError
exception when executing the test case in Listing~\ref{list:set}. However, the expected output is ``1,2,3,0,0'' per ECMA-262. The root
cause of this bug is because JSC does not convert the String object \texttt{e} at line 2 to an Array object to be used at line 4.
\SystemName can generate a bug-exposing test case by exploiting the ECMA-262 rule for \texttt{\%TypedArray\%.prototype.set}. This bug is
confirmed and fixed by JSC developers. This test case also triggered a similar bug in Graaljs.

\begin{lstlisting}[basicstyle=\footnotesize, frame=lines, caption={JSC prior to v261782 throws a TypeError while the code is correct per
ECMA-262.},upquote=true,captionpos=b, label={list:set}]
var foo = function(){
  var e = '123';
  A = new Uint8Array(5);
  A.set(e);
  print(A);
};
foo();
\end{lstlisting}

\vspace{-2.5mm} \cparagraph{Array allocation bug.} Listing~\ref{list:array} gives a bug-exposing test case generated by \SystemName. When
setting the object property at line 4, QuickJS appended the right-hand-side value to the end of the \texttt{obj} array (i.e., leading to an
array of 4 elements rather than setting the object property). As a result, when running this code with QuickJS, the program gives an
erroneous output of ``1,2,5,10$\backslash$n undefined''. This bug will only be triggered if the \texttt{property} is set to
\texttt{true}. By utilizing ECMA-262 to generate the test data, \SystemName is able to expose this implementation bug. This bug was
verified by the developer on the same day of submitting the bug report.

\begin{lstlisting}[basicstyle=\footnotesize, frame=lines, caption={QuickJS incorrectly appends the property value (line 4) as a new element
of the array object.},upquote=true, captionpos=b, label={list:array}]
var foo = function(){
  var property = true;
  var obj = [1,2,5];
  obj[property] = 10;
  print(obj);
  print(obj[property]);
};
foo();
\end{lstlisting}

\vspace{-2mm} \cparagraph{\texttt{eval} function.} When parsing the test case in Listing~\ref{list:eval}, a \JS compiler should throw a
SyntaxError exception. This is because according to ECMA-262, the \emph{for-loop} expression passed to \texttt{eval} should contain a loop
body or end up with a semicolon (i.e., an empty loop). However, ChakraCore allows this code to successfully compile and run, which is thus
a conformance bug. \SystemName generates this test case by exploiting the edge cases defined in ECMA-262. This bug was confirmed and fixed
by the ChakraCore developers, and our test case was also added into Test262 later.

\begin{lstlisting}[basicstyle=\footnotesize, frame=lines, caption={ChakraCore fails to throw a SyntaxError when parsing the expression
passed to \texttt{eval}.},upquote=true,captionpos=b,label={list:eval}]
var foo = function(cmd){
  eval(cmd);
  print("Run Here 1");
};
var str = "for(;false;)";
foo(str);
\end{lstlisting}

\vspace{-2mm} \cparagraph{\texttt{String.prototype.split}.} Listing~\ref{list:split} shows a test case that triggered a conformance bug of
JerryScript. The program applies a regular expression to split a String object starting with the capital letter `A'. In this case, the
program should produce ``anA'' as an output as the string does not match the regular expression. JerryScript yields ``an'' as the output
due to the incorrect implementation of it regular expression parsing engine. This test program was automatically generated by our GPT-2
code synthesizer by learning from open-source \JS programs, and it was also added to Test262.

\begin{lstlisting}[basicstyle=\footnotesize, frame=lines, caption={This test program exposes a bug of the JerryScript regular expression
parser.},upquote=true,captionpos=b,label={list:split}]
var foo = function(){
  var a = "anA".split(/^A/);
  print(a);
};
foo();
\end{lstlisting}

\cparagraph{Crash.} QuickJS crashed when compiling the test case in Listing~\ref{list:crash} when an empty string invokes
\texttt{normalize()}. \SystemName generates this test data by learning from open-source \JS code.  It was confirmed by the QuickJS
developers that this is a memory safety issue that could be exploitable to run arbitrary code. Hence, this bug was quickly fixed.

\begin{lstlisting}[basicstyle=\footnotesize, frame=lines, caption={Test case that leads to a QuickJS compilation
crash.},upquote=true,captionpos=b,label={list:crash}]
var foo = function(str){
  str.normalize(true);
};
var parameter = "";
foo(parameter);
\end{lstlisting}

\begin{figure}[t!]
    \centering
    \includegraphics[width=0.45\textwidth]{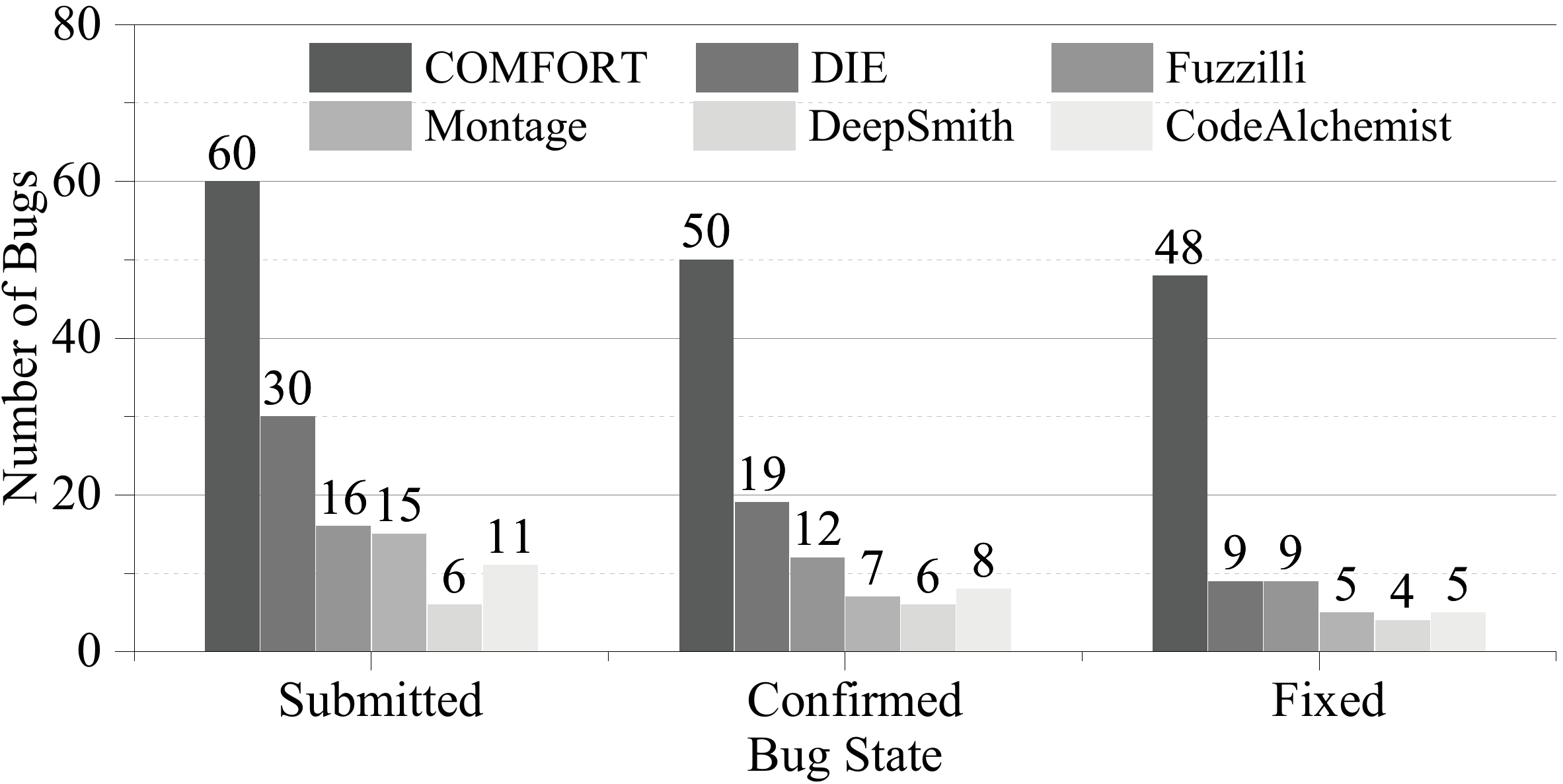}
    \caption{The number of bugs found by different fuzzers during 72 hours automated testing runs and a 3-month bug confirmation and fixing window.}
    \label{fig:comparision_bugs}
\end{figure}

\subsection{Compare to Prior Compiler Fuzzers\label{sec:comparison}}
We compare \SystemName against five compiler
fuzzers~\cite{cummins2018compiler,han2019codealchemist,son2020montage,gross2018fuzzil,park2020fuzzing}, of which four were specifically
designed for fuzzing \JS compilers~\cite{han2019codealchemist,son2020montage,gross2018fuzzil,park2020fuzzing}. We consider the capability
in exposing bugs (Sections \ref{sec:bec} and \ref{sec:otherc}) and the quality of the generated test cases (Section \ref{sec:tcg}).

\subsubsection{Bug exposing capability\label{sec:bec}}
In this experiment, we test each \JS Testbed for 72 hours of consecutive testing runs, using test cases generated by different fuzzers. An
average of 20k offline generated, syntactically correct test cases were evaluated on each Testbed. Here, the testing time refers to the
engine execution runtime (by excluding test case generation time).  To provide fair comparison, for mutation-based fuzzers, we use the seed
programs provided in the source publications for test case generation; and we train DeepSmith~\cite{cummins2018compiler} using the same
training \JS corpus as \SystemName.  In this experiment, we leave at least three months for the relevant \JS engine developers to confirm
and fix a submitted bug. Based on our experience, an important bug is usually confirmed by the developers within two weeks and gets fixed
within three months. Test runs were conducted in May 2020, and we ran the bug confirmation and fixing window until October 2020.  We
 exclude Nashorn in this experiment as it was no longer under active development since June 2020.

As can be seen from Figure \ref{fig:comparision_bugs}, \SystemName discovered more distinct bugs than any other individual fuzzer during
the testing period of this experiment. In 200 hours of testing, \SystemName discovered a total of 60 unique bugs across all
Testbeds. It helped the vendors fixed over 95\% of the confirmed bugs in the 3-month time frame. These numbers are more than the sum of fixed bugs
discovered by other fuzzers.
DeepSmith, the most closely related
DNN-based generative based fuzzer,  discovered a total of six bugs from four tested \JS engines. By contrast, \SystemName discovered bugs in all tested engines.  Furthermore, \SystemName alone discovered
31 confirmed bugs that were not uncovered by other fuzzers. By comparison, a total of 29 confirmed bugs discovered by five other fuzzers all togehter were not exposed
by \SystemName during the test runs.
This experiment shows that \SystemName is effective in uncovering \JS conformance bugs.

\subsubsection{Test cases generated by other fuzzers\label{sec:otherc}} We now discuss some bug-exposing test cases
generated by other fuzzers, which were not covered by \SystemName-generated test cases during this experiment.

 \cparagraph{CodeAlChemist.} The CodeAlChemist test case in Listing~\ref{list:u4} exposes a conformance bug of Rhino. For this case, a
TypeError exception should be thrown because a \texttt{null} argument is passed to \texttt{String.prototype.big.call}. \SystemName does not
generate such a case because none of the training \JS programs used to train our program generator uses \texttt{String.prototype.big.call}
and hence our program generator does not learn to generate test cases using this API.

\begin{lstlisting}[basicstyle=\footnotesize,caption={CodeAlChemist generated  test
case.},upquote=true,frame=lines,captionpos=b,label={list:u4}]
var v0 = (function(){
   print(String.prototype.big.call(null));
});
v0();
\end{lstlisting}

\cparagraph{Fuzzilli.} For the Fuzzilli generated test case in Listing~\ref{list:u6}, Rhino crashed when executing the seal function at
line 3. This is an implementation error that is not defined as a \ecma boundary case. Thus, \SystemName does not generate a similar test
case.

\begin{lstlisting}[basicstyle=\footnotesize,caption={Fuzzilli generated test
case.},upquote=true,frame=lines,captionpos=b,label={list:u6}]
function main(){
   var v2 = new String(2477);
   var v4 = Object.seal(v2);
}
main();
\end{lstlisting}

\cparagraph{DIE.} The test case generated by DIE in Listing~\ref{list:u3} exposes a bug of Rhino and JerryScript. At lines 2-5, the \JS
code set the \texttt{lastIndex} property of the \texttt{regexp5} object (defined at line 1) to be \emph{non-writable}. The program later at
line 5  compiles the regular expression,  which effectively will set the \texttt{lastindex} to 0 per the \ecma standard. For this case, a
TypeError exception should be thrown when executing the statement at line 5, but Rhino and JerryScript permit such a change (which thus is
a bug). This \ecma definition was given in the natural language form and hence is not captured by our \ecma parser (Section
\ref{sec:rulex}). \SystemName was able to generate test case to trigger a similar type of bug in Listing \ref {list:defineProperity}
because that definition is written in the pseudo-code form.

\begin{lstlisting}[basicstyle=\footnotesize,caption={DIE generated  test
case.},upquote=true,frame=lines,captionpos=b,label={list:u3}]
var regexp5 = new RegExp(/abc/);
Object.defineProperty(regexp5, "lastIndex", {
   value: "\w?\B", writable: false 
});
regex5 = regexp5.compile("def");
print(regexp5.lastIndex);
\end{lstlisting}

\cparagraph{Montage.} The test code generated by Montage in Listing~\ref{list:u5} is an \emph{immediately invoked function expression}
because the function at line 1 and the variable name at line 2 are both named \texttt{v1}. Hermes and Rhino produced an output of
``false$\backslash$n number'', while other \JS engines output ``true$\backslash$n function''. This is an undefined behavior in
\ecma and hence is not exposed by \SystemName. After submitting the report, the developers of Hermes and Rhino decided to label it as a
bug.

\begin{lstlisting}[basicstyle=\footnotesize,caption={Montage generated  test
case.},upquote=true,captionpos=b,frame=lines,label={list:u5}]
(function v1(){
   v1 = 20;
   print(v1 !== 20);
   print(typeof v1);
}());
\end{lstlisting}

\subsubsection{Quality of test cases generation\label{sec:tcg}}
To evaluate the quality of the generated test cases, we consider two metrics:

\cparagraph{Syntax passing rate.} This quantifies the ratio of the generated \JS programs that are syntactically valid (judging by JSHint -
a static \JS parser). We ask each fuzzer to generate 10,000 \JS programs and compute the passing rate.

\cparagraph{Coverage.} We consider three widely-used coverage metrics of the generated test
cases~\cite{klees2018evaluating,liu2019deepfuzz}: statement coverage, function coverage, and branch coverage. The three metrics
respectively measure the average ratio of statements, functions and branches of a \emph{test \JS program} that get executed during the test run. For
fair comparison, we randomly select 9,000 syntactically valid test programs generated by each fuzzer to compute the coverage. We use
\texttt{Istanbul}~\cite{istanbul}, a \JS code coverage tool, to collect the information.

\begin{figure}[t!]
    \centering
    \includegraphics[width=0.45\textwidth]{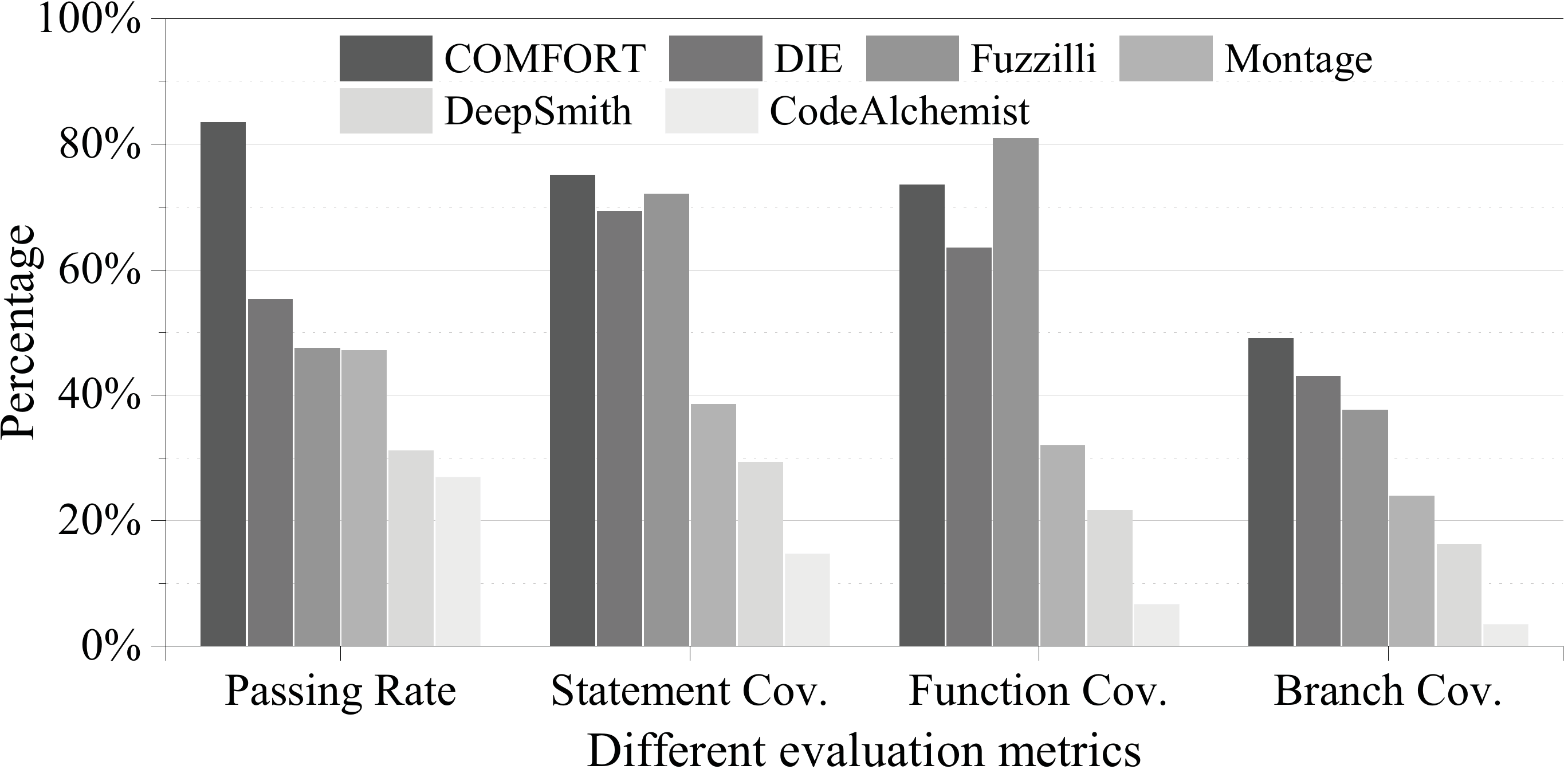}
    \caption{Compare to other fuzzers, \SystemName generates larger and more syntactically corrected test programs with a higher coverage of \JS APIs and branches. }
    \label{fig:quality}
\end{figure}

\cparagraph{Results.} As can be seen from Figure~\ref{fig:quality}, \SystemName gives a passing rate of 80\%, an improvement over the less
than 60\% passing rate given by all alternative methods. Note that among the syntactically correct test cases generated by \SystemName,
about 18\% of them triggered a runtime exception during testing due to semantic errors. \SystemName also gives the best statement and
branch coverages by ensuring most of the branches of a test program will be executed. Fuzzilli gives the best overall function coverage
through hand-crafted function generation rules using seed programs, but it gives lower statement and branch coverage rates compared to
\SystemName. While the test cases generated by Fuzzilli cover more functions than \SystemName, many of the statements and branches do not
get executed during execution.

\subsection{Bug Importance} It is reasonable to ask that if the bugs we found matter. It is not easy to provide a quantified answer to
this question. Here, we follow the discussion from the source publication of CSmith \cite{yang2011finding}. Some of our reported bugs have
been independently discovered and reported by application developers. This suggests we indeed report bugs that occurred in real-world
applications. As of November 2020, 21 of the test cases generated by \SystemName have been added to the ECMA-262 official test suite
because they triggered bugs in mainstream \JS engines, of which 3 of the \SystemName-generated test cases were submitted by the relevant
\JS compiler vendor. This shows that the standard body and \JS compiler developers appreciate the importance of the bugs triggered by
\SystemName. Finally, most of our reported bugs have been confirmed and fixed by the developers, illustrating their relevance and
importance (as it often takes substantial effort to fix a compiler defect); and the development teams of four tested \JS engines marked a
total of 25 of our bugs as a release-blocking priority for functional bugs. Furthermore, eight \SystemName-exposing bugs were welcomed and
quickly fixed by the compiler developers in less than 48 hours.

\subsection{Discussions and Future Work}

Our study has been in the context of \JS by utilizing the well-structured ECMA-262 specification to generate test data. We stress that
\SystemName is not designed to replace existing fuzzers. Instead, our experiment shows that \SystemName can provide useful complementary
test cases to discover conformance bugs by exploiting the  pseudo-code-like definitions in \ecma. There are still some definitions were
given in the natural language form in ECMA-262. These are not covered by \SystemName, and may still require expert involvement to
understand to manually create test-case generation rules or write test cases to cover.

Like most supervised learning methods, our program synthesizer can suffer from insufficient training data. In which case, we expect the
time in generating syntactically valid programs would remain largely unchanged, but the generated programs will be less diverse and,
therefore, will affect test coverage and the ability for exposing bugs. However, we believe this is not an issue for popular programming
languages like \JS with a large open-source code base.

 Our approach could be
applied to other languages.  Doing so would require the API semantics and expected behavior to be described in the pseudo-code form like
 ECMA-262. Our current implementation cannot exploit language specification written in free-form natural languages like the C standard. We
view this as an exciting open challenge: can we transfer the decades' research in the natural language comprehension to translate language
specifications to a form that can be exploited by a fuzzer? We have showcased that Transformer-like natural language processing models
(GPT-2 used in this work) can be useful in generating validate test programs. Our language model can generate syntax correct test programs
with a higher success rate than the state-of-the-art DL-based program generator. This approach is readily transferable to many other code
generation tasks as the language model infers the language syntax and semantics directly from the training corpus.  Like DeepSmith and many
other \JS fuzzers, \SystemName does not generate floating-point test programs. However, methods for testing floating-point programs
\cite{liew2019just} are orthogonal to our work.

\section{Related Work}
Random test case generation, in general, falls into two categories: program generation and program mutation. The former generates a test
case from scratch while the later modifies existing test cases to expose anomalous behavior. 

\cparagraph{Program generation.} Program generation often relies on stochastic context-free grammars. The fuzzer takes a grammar describing
the syntax of the language being tested to produce syntactically valid programs whose various expressions conform to a given probability
distribution of the grammar's productions. The Mozilla jsfunfuzz~\cite{funfuzz} was the first publicly available \JS fuzzer. It uses a set
of hand-crafted generation rules based on the \JS grammars to generate test cases. Similarly, Domato~\cite{domato} employs pre-defined
templates to generate random web programs (include \JS code) to test web browsers. Other works  also use customized grammar-based
constraint rules to generate test cases  \cite{majumdar2007directed, godefroid2008grammar, bielik2016phog, gopinath2018sample,
le2019saffron, bastani2017synthesizing, mathis2019parser}. PHOG~\cite{bielik2016phog} can generate programs with rich contexts features
with probabilistic context-free grammars. CSmith \cite{yang2011finding} is an effective program generator for generating C test programs
using pre-defined grammar rules. Subsequent generators influenced by CSmith, like CLSmith~\cite{clsmith}, have extended random program
generation to other programming languages. These approaches all require expert involvement and significant engineering effort to create the
generation rules, grammars or templates. \SystemName builds upon these past foundations by combining random program generation and language
standard-guided test data generation to generate test cases for \JS engines. \SystemName reduces human involvement by leveraging deep
learning to learn the language syntax and semantics from real-world code.

\cparagraph{Program mutation.} Mutation-based fuzzing modifies a set of seed programs to generate test cases~\cite{sutton2007fuzzing,
holler2012fuzzing,woo2013scheduling,cha2015program}. Equivalence modulo input testing is a representative mutation-based test case
generation method \cite{pldi14}.
Langfuzz~\cite{holler2012fuzzing} mutates test cases by inserting code segments that previously exposed bugs. SYMFUZZ~\cite{cha2015program}
utilizes the white-box symbolic analysis to search for useful program inputs by mutating a seed program and a seed input together.
IFuzzer~\cite{veggalam2016ifuzzer} uses evolutionary algorithms to generate unusual input code fragments. AFL~\cite{afl} and its subsequent
works~\cite{bohme2017coverage,
rebert2014optimizing,guo2013gramfuzz,woo2013scheduling,lemieux2018fairfuzz,mathis2020learning}  mutate
the seed program to improve the test run coverage. Such techniques can be useful for \SystemName in improving the test run coverage.
DIE~\cite{park2020fuzzing} and CodeAlChemist~\cite{han2019codealchemist} are two mutation-based fuzzers for \JS. DIE employs a stochastic
process to preserve properties which are likely to expose bugs across mutations. CodeAlChemist breaks the seed programs into fragments and
uses the fragments to assemble new \JS test cases. By contrast, \SystemName is a generative approach that does not require access to a set
of seed program inputs. It is a pure black-box approach, requiring no source code, seed test programs, or other knowledge of the target
compiler. AutoTest \cite{meyer2009programs} is a contract-based random testing tool. It uses fuzzing techniques to test the conformance of
the Eiffel program against the given contracts. It can expose bugs in the runtime system and the associated library implementation. By
contrast, \SystemName is designed to test compiler-specific implementations of APIs, non-API features, and compiler components like the
parser, code generator, and optimizer. Nonetheless, extending \SystemName to mutate bug-exposing test cases could be valuable.

\cparagraph{Deep learning for compiler testing.} Recently, deep learning (DL) models have been used for code modeling
\cite{code2vec,code2seq,cummins2017end,ye2020deep,wang2020combining}, random program
generation~\cite{cummins2017synthesizing,sun2019grammar} and input fuzzing \cite{godefroid2017learn}. DeepSmith ~\cite{cummins2018compiler}
and DeepFuzz~\cite{liu2019deepfuzz} are two closely related works. Both approaches use the recurrent neural network (RNN), e.g., LSTM, to
generate test programs. Montage~\cite{son2020montage} is a mutational \JS fuzzer. It produces test cases by replacing the code snippets of
the seed program's abstract syntax tree with a new code fragment generated by a LSTM model. Due to the limitation of RNN in capturing the
long-term dependence of source code, they often generate many syntactically invalid programs that are rejected by the \JS engines in the
parsing stage. Our approach replaces the RNN generation model with a more advanced neural network, significantly improving the number of
syntactically valid generated programs. None of the existing DL-based fuzzers has exploited the language specification to assist test case
generation. \SystemName is the first in doing so.

\cparagraph{Conformance testing for \JS.} Our work was conducting concurrently with JEST~\cite{park2021jest} that also leverages
differential testing and \JS specifications to expose conformance bugs in \JS compiler implementations. JEST utilizes
JISET~\cite{park2020jiset}, a \JS semantics extraction tool, to extract specification from the \ecma document. Unlike \SystemName,  JEST is
a program mutation approach that relies on a set of seed programs to create the \JS test cases. \SystemName thus has the advantages of not
depending on the quality of the seed programs.  Nonetheless, it would be interesting to extend \SystemName to use the semantics extracted
by JISET to perform differential testing.

\section{Conclusions}
We have presented \SystemName, a novel compiler fuzzing framework for testing standard conformance bugs for \JS compilers. \SystemName
leverages the recent advance in the deep-learning-based language model to automatically generate test \JS programs without hand-crafted
grammar or generation rules. It then augments the test programs with code and variables derived from the \JS standard rules as test data to
expose \JS compiler bugs. We evaluate \SystemName by applying it to test ten mainstream \JS engines. In 200 hours of automated concurrent
test runs, we found bugs in all the  \JS compilers we tested. At the time of submission, \SystemName has discovered \VBugs unique,
confirmed bugs, of which \FBugs bugs have been fixed by the developers, and \TSC of the \SystemName-generated test cases have been added
into the official \JS conformance test suite. Our work showcases a new way to leverage a structured language specification to produce
bug-exposing test data to detect standard conformance bugs in compiler implementations, opening up an exciting research avenue.

\section*{Acknowledgements}
We thank our shepherd, Jos\'{e} Fragoso Santos, and the anonymous PLDI reviewers for their valuable feedback. We also thank Yang Tian,
Houyou Yao, Xing Qu, Wen Yi and Yuan Wang for their help in verifying and submitting bug reports. This work was supported in part by the
National Natural Science Foundation of China (NSFC) under grant agreements 61972314, 61872294 and 61902170, the International Cooperation Project of
Shaanxi Province under grant agreements 2021KW-04, 2020KWZ-013 and 2021KW-15,  an Ant Financial Science funded project and an Alibaba
Innovative Research Programme grant.

\newpage

\bibliographystyle{ACM-Reference-Format}
\balance
\bibliography{ref}

\end{document}